%

\ifx\mnmacrosloaded\undefined \input mn\fi

%

\newif\ifAMStwofonts

\ifCUPmtplainloaded \else
  \NewTextAlphabet{textbfit} {cmbxti10} {}
  \NewTextAlphabet{textbfss} {cmssbx10} {}
  \NewMathAlphabet{mathbfit} {cmbxti10} {} 
  \NewMathAlphabet{mathbfss} {cmssbx10} {} 
  \ifAMStwofonts
    \NewSymbolFont{upmath} {eurm10}
    \NewSymbolFont{AMSa} {msam10}
    \NewMathSymbol{\upi}     {0}{upmath}{19}
    \NewMathSymbol{\umu}     {0}{upmath}{16}
    \NewMathSymbol{\upartial}{0}{upmath}{40}
    \NewMathSymbol{\leqslant}{3}{AMSa}{36}
    \NewMathSymbol{\geqslant}{3}{AMSa}{3E}

     \let\le=\leqslant
     \let\ge=\geqslant
  \else
    \def\umu{\mu}
    \def\upi{\pi}
    \def\upartial{\partial}
  \fi
\fi


\pageoffset{0pc}{0pc}

\loadboldmathnames



\onecolumn        

\begintopmatter  

\title{On a matrix method for the study of small perturbations in galaxies}
\author{Peter O. Vandervoort}
\affiliation{Department of Astronomy and Astrophysics, The University of
Chicago, 5640 Ellis Avenue, Chicago, IL 60637, USA}

\shortauthor{P. O. Vandervoort}
\shorttitle{Small perturbations in galaxies}



\abstract {A matrix method is formulated in a Lagrangian
representation for the solution of the characteristic
value problem governing modes of oscillation and
instability in a collisionless stellar system.  The underlying
perturbation equations govern the Lagrangian displacement of
a star in the six-dimensional phase space.  This matrix method
has a basis in a variational principle.  The method is
developed in detail for radial oscillations
of a spherical system.  The basis vectors required for the
representation of the Lagrangian displacement in this
case are derived from solutions of
the Lagrangian perturbation equations for radial perturbations
in a homogeneous sphere.  The basis vectors are made divergence
free in the six-dimensional phase space in accordance with
the requirement of Liouville\rq{s} theorem that the
flow of the system in the phase
space must be incompressible.  The basis vectors are made
orthogonal with respect to a properly chosen set of adjoint
vectors with the aid of a Gram-Schmidt procedure.  Some
basis vectors are null vectors in the sense that their inner
products with their own adjoint vectors vanish.  The
characteristic frequencies of the lowest radial modes are calculated
in several approximations for members of a family of spherical
models which span a wide range of central concentrations.  The
present formulation of the matrix method can be generalized for
nonradial modes in spherical systems and for modes in axisymmetric
systems.}


\maketitle  

\section{Introduction}

Considerations of stability and instability impose certain constraints
on the equilibrium and evolution of a galaxy.
Most galaxies observed in nature are presumed to be in
stable states of equilibrium, or nearly so, inasmuch as they would not
be expected to survive in unstable states long enough to be observed.
If the processes of formation or the subsequent processes of evolution
bring a galaxy into an unstable state, then the onset and
development of that instability would act as dynamical mechanisms of
evolution.  Consequences of the instability could be manifest in a
later equilibrium state of such a galaxy.

Examples of instabilities in galaxies are well known.  According to
Toomre (1964), an axisymmetric disk would be locally unstable with
respect to axisymmetric perturbations unless the dispersion of the
peculiar velocities of the stars exceeded a certain critical value.
The conjecture of Ostriker \& Peebles (1973) that excessive rotation
would make an axisymmetric galaxy unstable with respect to a deformation
into a bar has become a central organizing principle in efforts to
understand the dynamics of spiral galaxies.  A third example, originally
suggested by Antonov (1973) and demonstrated for the first
time by Polyachenko \& Shukhman (1981) and Polyachenko (1981), is the
radial orbit instability in which the anisotropy of the velocity
distribution makes a spherically symmetric stellar system unstable
with respect to an axisymmetric deformation.
Finally, certain firehose instabilities can deform the equatorial
plane of an axisymmetric
stellar system if the system is sufficiently flattened, and those
instabilities might, according to Fridman \& Polyacheko
(1984; see Chapter X), explain the absence of elliptical galaxies
flatter than type E7 on the Hubble sequence.

It is beyond the scope of this introduction to present an
extensive review of the history of investigations of such instabilities.
An introduction to the subject is contained in the textbook of Binney
\& Tremaine (1987), and the subject is treated extensively in the
monographs of Fridman \& Polyachenko (1984) and Palmer (1994).
Reviews by Merritt (1987, 1990) and Polyachenko (1987) are useful
supplements to the monographs.  Accounts of subsequent investigations
(e.g., Merritt \& Sellwood 1994; Sellwood \& Merritt 1994; Robijn 1995;
Sellwood \& Valluri 1997) include references to more recent developments.

For the work described in this paper, the most relevant part of the
literature on small perturbations in galaxies concerns the matrix
method formulated by Kalnajs (1977) for the solution of the
characteristic value problem governing normal modes of oscillation
and instability in a collisionless stellar system.
Originally formulated for the study of instabilities
in axisymmetric disks, that method has subsequently been used
in several investigations of the radial orbit instability in spherical
systems (Polyachenko \& Shukhman 1981; Fridman \& Polyachenko 1984;
Palmer \& Papaloizou 1987; Weinberg 1989,1991a; Saha 1991,
Bertin et al. 1994).
More recently, the method has been extended to the investigation
of small perturbations in axisymmetric systems (Robijn 1995).
The matrix method of Kalnajs is based on an Eulerian representation
of the perturbations.

A Lagrangian representation of the perturbations provides an
alternative framework for the investigation of the oscillations and
the stability of stellar systems (Vandervoort 1983, 1989, 1991).
This paper describes a matrix method for the study of normal modes in
the Lagrangian representation.  The elements of Lagrangian perturbation
theory that provide the foundation for the method are
reviewed in Section 2, and the method is formulated
in Section 3.  For the sake of definiteness and simplicity in this first
presentation of the new matrix method, we concentrate on an application of
the method to the study of radial modes of oscillation
in a spherically symmetric galaxy.  In Section 4 a procedure is described for
the construction of basis vectors which can be used for the representation
of such perturbations.  The reduction and solution of the characteristic 
value problem in the matrix representation is described
in Section 5, and Section 6 contains examples of solutions for modes
which illustrate and test the method.  A connection between the discrete
and continuous spectra of modes in a stellar system is described in
Section 7.  Important technical details are described in a series of
appendices.

As this paper was being written, substantial progress was made in the
generalization of the present formulation of the matrix method for
a wider range of applications.  The most important details of such
generalizations have already been worked out for non-radial modes of
oscillation in spherical systems and for general oscillations of
axisymmetric systems.  The construction of basis vectors for the
representation of non-radial modes in spheres follows the procedure
described in Section 4 very closely.  It consequently appears that
much of the analysis described in this paper for radial oscillations
in spheres can be readily extended
to the study of non-radial oscillations in spheres.  A similar
construction of basis vectors for the representation of perturbations
in axisymmetric systems can be achieved with the aid of a formulation
based on an earlier investigation of the lowest modes of oscillation of a
homogeneous spheroid (Vandervoort 1991).  Finally, preliminary investigation
suggests that this approach to the construction of basis vectors
can be extended to nonrotating triaxial systems.  The author plans
next to apply these generalizations, in a continuation of
the present work, to the radial orbit instability in spherical systems
and to firehose instabilities in rotating axisymmetric systems.

\section{Lagrangian perturbation theory}

Small perturbations of a collisionless stellar system are described
in the Lagrangian representation in terms of the Lagrangian
displacement $(\Delta { \bmath{x} }, \Delta { \bmath{v} } )$
of a star (Vandervoort 1983).  Thus, if the position and velocity
of the star at time $t$ would have been ${ \bmath{x} }$ and
${ \bmath{v} }$, respectively, in the unperturbed system, then the
position and velocity of the star at that time are, by definition,
$ { \bmath{x} } + \Delta { \bmath{x} } $ and
$ { \bmath{v} } + \Delta { \bmath{v} } $, respectively, in the perturbed
system.

For an infinitesimal perturbation, $ \Delta { \bmath{x} } $ and
$ \Delta { \bmath{v} } $ are considered to be functions of
$ {\bmath{x} } $, $ { \bmath{v} } $, and $t$, and the equations of motion
for the perturbation are
$$
{ { { {\rm d} \Delta { \bmath{x} } } \over { {\rm d}t } } = \Delta { \bmath{v} } }
\quad \hbox{ and } \quad
{ { { {\rm d} \Delta { \bmath{v} } } \over { {\rm d}t } } =
\Delta { \bmath{a} } \{ \Delta { \bmath{x} } \} } , \eqno\stepeq
$$
where
$$
{ {\rm d} \over { {\rm d}t } } =   { \upartial \over { \upartial t } } +
{ \bmath{v} } \cdot { \upartial \over { \upartial { \bmath{x} } } } -
{ { \upartial V_0 } \over { \upartial { \bmath{x} } } } \cdot 
{ \upartial \over { \upartial { \bmath{v} } } } \eqno\stepeq
$$
is the total time derivative along the unperturbed 
trajectory of the star, and
$$
\Delta { \bmath{a} } \{ \Delta { \bmath{x} } \} =
- \left( \Delta { \bmath{x} } \cdot { \upartial \over { \upartial { \bmath{x} } } }
\right) { { \upartial V_0 } \over { \upartial { \bmath{x} } } }
 - { \upartial \over { \upartial { \bmath{x} } } } \left[ { G m_* 
{ \upartial \over { \upartial { \bmath{x} } } } \cdot \int_\Omega
{ { { \Delta { \bmath{x} } ({\bmath{x} }^\prime , { \bmath{v} }^\prime ,t)
f_0 ({\bmath{x} }^\prime , { \bmath{v} }^\prime ) } \over
{ | { \bmath{x} }^\prime - { \bmath{x} } | } }
{\rm d} { \bmath{x} }^\prime {\rm d} { \bmath{v} }^\prime } } \right] 
\eqno\stepeq
$$
is the Lagrangian perturbation of the acceleration of the star.
Here the distribution function $ f_0 ({\bmath{x} } , { \bmath{v} } ) $
represents the unperturbed density of stars in the six-dimensional phase 
space of a single star, and $V_0 ( { \bmath{x} }) $  represents the 
gravitational potential in the unperturbed system.  Also, $G$ 
denotes the constant of gravitation, and $m_*$ denotes the 
mass of a single star.  The second term on the right-hand 
side of equation (3) represents the Eulerian perturbation of 
the gravitational acceleration of a star, and the integration 
there extends over the region $\Omega$ of the phase space that is 
accessible to stars in the unperturbed system.

Equations (1)-(3) are a homogeneous system of linear integro-differential
equations governing an infinitesimal perturbations of a stellar
system.  In Vandervoort (1989, hereafter LM), these equations have
been reduced to a characteristic-value problem governing the normal
modes of oscillation and instability of a system.
The subsections that follow contain a review of the results
derived in LM that are essential for the present formulation of
a matrix method.  

\subsection{The characteristic value problem}

For normal modes of oscillation having a time dependence of the
form $ \exp (- {\rm i} \omega t ) $, where the characteristic frequency
$ \omega $ is a constant, we represent the Lagrangian displacement
as a six-vector
$$
{ { \boldeta } = \left(\matrix{ \Delta { \bmath{x} } \cr
        \Delta { \bmath{v} }\cr }\right) }  \eqno\stepeq
$$
in the six-dimensional phase space, and we reduce equations (1)
to the form
$$
\omega {\boldeta} = P {\boldeta} .  \eqno\stepeq
$$
Here $P$ is the matrix of operators
$$
{  P  =  \left(\matrix{-D&{\rm i}\cr
           {\rm i} \Delta { \bmath{a} \{ \} }&-D\cr}\right)    } ,
\eqno\stepeq 
$$
where
$$
{ D } =   {\rm i} \left({
{ \bmath{v} } \cdot { \upartial \over { \upartial { \bmath{x} } } } -
{ { \upartial V_0 } \over { \upartial { \bmath{x} } } } \cdot 
{ \upartial \over { \upartial { \bmath{v} } } } } \right) .
\eqno\stepeq
$$

We identify $\boldeta$ as an element of a Hilbert space defined by
letting the inner product of two six-vectors
$$
{\bmath{g} ({\bmath{x} } , { \bmath{v} } )
= \left( \matrix{ \bmath{g}_1 ({\bmath{x} } , { \bmath{v} } ) \cr
        \bmath{g}_2 ({\bmath{x} } , { \bmath{v} } ) \cr }\right) }
\quad \hbox{ and } \quad
{ \bmath{h} ({\bmath{x} } , { \bmath{v} } )
= \left(\matrix{ \bmath{h}_1 ({\bmath{x} } , { \bmath{v} } ) \cr
        \bmath{h}_2 ({\bmath{x} } , { \bmath{v} } ) \cr }\right) } ,
\eqno\stepeq
$$
say, where $\bmath{g}_1$, $\bmath{g}_2$, $\bmath{h}_1$, and
$\bmath{h}_2$ are three-vectors in the six-dimensional
phase space, be
$$
{ \langle { \bmath{g}, \bmath{h} } \rangle}_2
= m_* \int_{\Omega} {\bmath{g}^{*} ({\bmath{x} } , { \bmath{v} } ) }
\cdot {\bmath{h} ({\bmath{x} } , { \bmath{v} } ) } 
f_0 ({\bmath{x} } , { \bmath{v} } )
{\rm d} {\bmath{x} } {\rm d} {\bmath{v} } ,  \eqno\stepeq
$$
where the asterisk, written as a superscript, signifies complex
conjugation here and in what follows.  In the notation adopted in LM,
the subscript 2 appears on the left-hand side of equation (9) in
order to distinguish the inner products of six-vectors from the
inner products of three-vectors.  The present definition of the
inner product of two six-vectors differs slightly from the
definition in LM; we now include a factor $m_*$ which was previously
omitted.  It is to be emphasized here that $ \bmath{g}^{*} \cdot \bmath{h}$
is the scalar product of two six-vector functions in the six-dimensional
phase space.

It is shown in the Appendix of LM that the Lagrangian displacement
satisfies a divergence condition of the form
$$
\bnabla_6 \cdot { \boldeta } \equiv
{ \upartial \over { \upartial { \bmath{x} } } } \cdot  \Delta { \bmath{x} }
+ { \upartial \over { \upartial { \bmath{v} } } } \cdot  \Delta { \bmath{v} }
= 0 .  \eqno\stepeq
$$
This condition implies, in accordance with Liouville's theorem, that
the perturbed flow of the stars in the phase space is incompressible.

\subsection{The adjoint problem}

The operator $P$ is not Hermitian in the Hilbert space of the
Lagrangian displacements.  Indeed, the operator adjoint to $P$ is
$$
{  P^A  =  \left(\matrix{-D&-{\rm i} \Delta { \bmath{a} \{ \} }\cr
           -{\rm i}&-D\cr}\right)    } ,  \eqno\stepeq
$$
for it is shown in LM that, for any two six-vectors
$\bzeta (\bmath{x} ,\bmath{v})$ and $\boldeta (\bmath{x} ,\bmath{v})$,
say, we have
$$
{ \langle { { \bzeta }, P { \boldeta } } \rangle}_2
= { \langle { P^A { \bzeta }, { \boldeta } } \rangle}_2 .  \eqno\stepeq
$$

Inasmuch as the characteristic value problem represented by equation (5)
is not self-adjoint, it must be solved in association with the solution
of the adjoint problem
$$
\omega^{*} {\boldeta^A} = P^A {\boldeta^A}  \eqno\stepeq
$$
governing vectors
$$
{ { \boldeta }^A = \left(\matrix{ \Delta { \bmath{v} }^A \cr
        \Delta { \bmath{x} }^A \cr }\right) }  \eqno\stepeq
$$
adjoint to the characteristic vectors ${\boldeta}$.  The notation
on the right-hand side of equation (14) is intended to emphasize that
we let the three-vectors $ \Delta { \bmath{v} }^A $ and
$ \Delta { \bmath{x} }^A $ have units of a velocity and a length,
respectively, so that the unit of an inner product such as
$ { \langle { { \bzeta }^A, { \boldeta } } \rangle}_2 $, say,
is well defined.

As is explained in Section IId of LM, the association of the
characteristic value problem with a distinct adjoint problem
introduces some complication, in general, into the solution
of the characteristic value problem.  However, as is further
explained in Section III of LM, the situation is greatly simplified
in cases in which the structure of the unperturbed system is
invariant with respect to the simultaneous application of
time reversal and reflection through a plane.  For the sake of
definiteness we express that invariance by writing
$$
f_0 (x_1,-x_2,x_3,-v_1,v_2,-v_3) = f_0 (x_1,x_2,x_3,v_1,v_2,v_3) ,
  \eqno\stepeq
$$
where we are letting $x_i$ and $v_i$ ($i=1,2,3)$ denote the
Cartesian components of $\bmath{x}$, and $\bmath{v}$, respectively,
and we are letting the plane of reflection be the $(x_1,x_3)$-plane.
In this paper, we formulate and study the matrix method for
systems having the symmetry described by equation (15).  Examples
of such systems include spherically symmetric systems, rotating
axisymmetric systems in which the unperturbed distribution function
$ f_0 (\bmath{x}, \bmath{v} )$ can be expressed in accordance
with the theorem of Jeans as a function of the energy integral and
the conserved component of the angular momentum of a star, and
systems (including axisymmetric systems) with triplanar symmetry
in which the gravitational potential $V_0 (\bmath{x})$ is of
the St{\"a}ckel form and $ f_0 (\bmath{x}, \bmath{v} )$ is
expressible in terms of three isolating integrals.

The simplification of the characteristic value problem for such
systems arises, because it is a consequence of equation(15) that
a characteristic vector $ \boldeta (\omega) $ and its adjoint
vector $ \boldeta^A (\omega^*) $ satisfy the relation
$$
{\boldeta}^A ( \omega^* ) = A {\boldeta} (\omega ) ,  \eqno\stepeq
$$
where $A$ is the matrix
$$
{  A  =  \left(\matrix{0&C\cr
           C&0\cr}\right)    } ,  \eqno\stepeq 
$$
and the operator $C$ transforms a three-vector function
$ \bmath{g} ( \bmath{x} , \bmath{v},t ) $ (say) in the phase space
in the manner
$$
{  \bmath{g} ( \bmath{x} , \bmath{v},t )
=  \left(\matrix{g_1 (x_1,x_2,x_3,v_1,v_2,v_3,t) \cr
                 g_2 (x_1,x_2,x_3,v_1,v_2,v_3,t) \cr
                 g_3 (x_1,x_2,x_3,v_1,v_2,v_3,t) \cr}\right)    }
\quad \rightarrow \quad
{  C \bmath{g} ( \bmath{x} , \bmath{v},t )
=  \left(\matrix{{g_1}^* (x_1,-x_2,x_3,-v_1,v_2,-v_3,-t) \cr
                 -{g_2}^* (x_1,-x_2,x_3,-v_1,v_2,-v_3,-t) \cr
                 {g_3}^* (x_1,-x_2,x_3,-v_1,v_2,-v_3,-t) \cr}\right)    } .
  \eqno\stepeq 
$$

The simultaneous application of time reversal,
reflection through the $ (x_1,x_3) $-plane, and complex conjugation
transforms $ \Delta \bmath{x} $ and $ \Delta \bmath{v} $ in the manner
$$
\Delta \bmath{x} (\bmath{x},\bmath{v},t) \rightarrow
C \Delta \bmath{x} (\bmath{x},\bmath{v},t)
\quad \hbox { and } \quad
\Delta \bmath{v} (\bmath{x},\bmath{v},t)
\rightarrow - C \Delta \bmath{v} (\bmath{x},\bmath{v},t) , \eqno\stepeq
$$
respectively.  The operator $C$ commutes with the operators $D$
and $ \Delta \bmath{a} \{ \} $ in the sense that
$$
C ( D \bmath{g} ) = D ( C \bmath{g} )
\quad \hbox { and } \quad
C(  \Delta \bmath{a} \{ \bmath{g} \} ) = \Delta \bmath{a} \{ C \bmath{g} \} ,
\eqno\stepeq
$$
where $ \bmath{g} ( \bmath{x} , \bmath{v},t ) $ is any three-vector function
in the six-dimensional phase space.

\section{Formulation of a Rayleigh-Ritz matrix method}

We turn now to the matrix method for the solution of the
characteristic value problem.  The method is formulated here
for normal modes whose frequencies form a discrete spectrum.

\subsection{A variational principle}

The basis for this matrix method is the following variational
principle, which was proved in Section IIIc of LM.  Let
$$
{ { \boldeta } = \left(\matrix{ \Delta { \bmath{x} } \cr
        \Delta { \bmath{v} }\cr }\right) }  \eqno\stepeq
$$
be an arbitrary six-vector in the six-dimensional phase space,
and let the value of $\omega$ be determined by the relation
$$
{ \langle { A { \boldeta } , (\omega - P ) { \boldeta } } 
\rangle}_2 = 0 .  \eqno\stepeq
$$
Let $\delta \omega$ be the change of $\omega$ that results when
$\boldeta$ is subjected to an arbitrary variation $\delta \boldeta$
in equation (22).  Then, under conditions in which
${ \langle { A { \boldeta } , { \boldeta } } \rangle}_2 \ne 0$,
a necessary and sufficient condition that $\delta \omega = 0$ is
that $\omega$ and $\boldeta$ satisfy the equation
$$
\omega {\boldeta} = P {\boldeta}  \eqno\stepeq
$$
and, accordingly, that $\boldeta$ is the characteristic vector
belonging to the characteristic frequency $\omega$.

\subsection{The matrix method}

Let the functions $ { \bbeta } ({\bf x},{\bf v}; k ) $
$(k = 1,2, \dots .)$ be elements of a complete set of basis
vectors in the Hilbert space of the Lagrangian displacements,
and let these basis vectors satisfy the orthogonality conditions
$$
{ \langle { A { \bbeta } ( j ), { \bbeta } ( k ) } \rangle}_2
= 0 \quad \hbox{ if } \quad j \ne k .  \eqno\stepeq
$$
In view of equation (10), we require that the basis vectors
also satisfy the divergence condition
$\bnabla_6 \cdot { \bbeta } (k) = 0$.
We represent the Lagrangian displacement as a superposition
of the basis vectors of the form
$$
{ \boldeta } = \sum_k a(k) { \bbeta } ( k ) ,  \eqno\stepeq
$$
where the coefficients $a(k)$ are constants.

In order to apply the variational principle, we identify the
right-hand side of equation (25) as a trial function for the Lagrangian
displacement in which the coefficients $a(k)$ are the variational
parameters.  Thus, we substitute from equation (25) for $\boldeta$
in equation (22), and we make use of the definition of the inner product
in equation (9), the definition of the operator $A$ in equations (17)
and (18), and the orthogonality relations in equation (24) in order
to simplify the resulting equation.  We next subject the resulting equation
to mutually independent, arbitrary variations $\delta a(k)$ of the
coefficients $a(k)$.  We require that the variation $ \delta \omega $
of the frequency $\omega$ vanish in the variational equation that
is obtained in this process.  As a consequence of the independence and
arbitrariness of the variations $\delta a(k)$, the variational
equation can be satisfied, in general, only if the frequency $\omega$
and the coefficients $a(k)$ satisfy the system of equations
$$
\omega { \langle { A { \bbeta } ( k ),{ \bbeta } ( k ) } 
\rangle}_2 a(k) 
=  \sum_j {1 \over 2} 
[{ \langle { A { \bbeta } ( j ),P { \bbeta } ( k ) } \rangle}_2
 + { \langle { A { \bbeta } ( k ),P { \bbeta } ( j ) } \rangle}_2 ]
 a(j) \qquad (k=1,2,\dots .) .   \eqno\stepeq
$$
This is the matrix form of the characteristic value problem
governing the normal modes of the system in the Lagrangian
representation.  Equations (26) are a homogeneous system of linear
equations in the coefficients $a(k)$.  In general, a solution for
those coefficients exists if and only if the determinant of
the secular matrix of the system vanishes.  That condition
provides the characteristic equation for the determination of
the frequencies $\omega$ of the modes.

One might have thought that the matrix representation of the
characteristic value problem could be derived directly from
equation (5).  However, it is not clear how, in such a direct
approach, one would know that the orthogonality relations must
be formulated in terms of inner products of the basis vectors
and their adjoint vectors, that the matrix of the operator $P$
is to be constructed as it is in terms of the basis vectors
and their adjoint vectors, or that the matrix of the operator
$P$ is to be made symmetric as it is on the right-hand side of
equation (26).  On the other hand, these conditions are imposed
automatically in the present derivation of the matrix
equations from the variational principle.  Thus the variational
principle would appear to be an essential foundation for the
matrix method.

We have deliberately retained the inner products
${ \langle { A { \bbeta } ( k ),{ \bbeta } ( k ) } \rangle}_2 $
of the basis vectors and their adjoint
vectors on the left-hand side of equation (26).  Ordinarily, we might
expect to normalize the basis vectors so that each of these
inner products is equal to unity.  However,
we have found in practice, as is explained in Section 4 below,
that the inner products of some basis vectors with
their own adjoint vectors can vanish.  In other words,
${ \langle { A { \bbeta } ( k ),{ \bbeta } ( j ) } \rangle}_2 $
is a diagonal matrix, in virtue of the orthogonality relations,
with some vanishing elements on the principal diagonal.  The
non-vanishing elements on the principal diagonal
may be normalized to unity in the usual way.

Equations (26) are an infinite system of equations.
In practice, we approximate the characteristic
value problem by expressing the Lagrangian displacement as a
superposition of a finite set of basis vectors $\bbeta (k)$
on the right-hand side of equation (25) and truncating the system
of equations (26) accordingly.  In the following sections, we
consider the solution of the characteristic value problem in
such an approximation.

\section{A set of basis vectors for the representation of
radial oscillations of a spherical galaxy}

In the remainder of this paper, we describe and illustrate the matrix
method with the aid of an application to radial modes of
oscillation of a spherical stellar system.  Consistently with the theorem of
Jeans, we let the distribution function of the unperturbed system be of the
form $f_0 (\bmath{x},\bmath{v})  = f_0 (E,L^2 )$, a function of
the energy $E$ and total angular momentum $L$ of a star.  Accordingly,
the distribution function $f_0 (\bmath{x},\bmath{v})$ is an even function of
the components of the velocity $\bmath{v}$.  This section is devoted
to a procedure for the construction of the basis vectors required
for the representation of radial perturbations in such a sphere.

The essential idea of the procedure is to construct a set of
basis vectors which would suffice for the exact solution of the
equations of motion governing the normal modes of oscillation of a
particular model of a stellar system.  The model chosen in the present
case is a homogeneous sphere.  Although the homogeneous sphere is an
unrealistic model of a galaxy, the resulting basis vectors appear to give good
results for the frequencies of radial modes even in quite centrally
concentrated systems (see Section 6 below).

For a homogeneous sphere of stars in which the unperturbed distribution
function $f_0 (\bmath{x},\bmath{v})  = f_0 (E,L^2 )$ is of a suitable
form, the radial modes of oscillation have Eulerian perturbations of
the gravitational potential which are polynomials in the square of
the radial coordinate $r = | \bmath{x} |$ (see Chapter III,
Section 4.1 of Fridman \& Polyachenko 1984).  The present
construction of the basis vectors is accordingly based on
solutions of the Lagrangian perturbation
equations for the response of a homogeneous sphere to an imposed
Eulerian perturbation of the gravitational potential which is a
polynomial in $r^2$.  The solution for the Lagrangian
displacement of the response is composed of six-vectors in the
phase space which are polynomials in the Cartesian components of
the position $ \bmath{x} $ and velocity $ \bmath{v} $ of a star.
With a slight generalization for applications to systems with inhomogeneous
density distributions, those \lq{six-vector polynomials}\rq\/ are almost the
basis vectors that we are seeking.  They are divergence free in the
phase space, but they are not orthogonal in the sense of equation (24).
The set of basis vectors finally adopted is constructed from the set of
vector polynomials with the aid of a Gram-Schmidt procedure for
satisfying the orthogonality conditions.  Like the vector polynomials,
the basis vectors are polynomials in the Cartesian components of
$ \bmath{x} $ and $ \bmath{v} $.

\subsection{Perturbations of a homogeneous sphere.}

For the purpose of investigating radial perturbations of a homogeneous
sphere, it is convenient to reduce the Lagrangian perturbation
equations as follows.  In the interior of the sphere, the unperturbed
gravitational potential is
$$
V_0 ( \bmath{x} ) = { { 2 \upi } \over 3 } G \rho_0 r^2
= {1 \over 2} {n_0}^2 r^2 , \eqno\stepeq
$$
apart from an additive constant, where $ \rho_0 $ is the density
in the sphere,
$$
{n_0}^2 = { { 4 \upi } \over 3 } G \rho_0 , \eqno\stepeq
$$
and $r = | \bmath{x} | $ is the radial coordinate of the point
$ \bmath{x} $ in a system of spherical polar coordinates with 
the origin at the centre of the sphere.  Stellar orbits in the
unperturbed system are simple harmonic oscillations with
frequency $ n_0 $.

In the expression for $ \Delta \bmath{a} \{ \Delta \bmath{x} \} $ in the
second of equations (1) (see also eq. [3]), we substitute from equation (27) for
$ V_0 ( \bmath{x} ) $ and we identify
$$
V_1( \bmath{x}, t ) =
{ G m_* { \upartial \over { \upartial { \bmath{x} } } } \cdot \int_\Omega
{ { { \Delta { \bmath{x} } ({\bmath{x} }^\prime , { \bmath{v} }^\prime ,t)
f_0 ({\bmath{x} }^\prime , { \bmath{v} }^\prime ) } \over
{ | { \bmath{x} }^\prime - { \bmath{x} } | } }
{\rm d} { \bmath{x} }^\prime {\rm d} { \bmath{v} }^\prime } } 
   \eqno\stepeq
$$
as the Eulerian perturbation of the gravitational potential (see eq.[14]
in Vandervoort[1983]).  We now eliminate $ \Delta { \bmath{v} } $ between
equations (1), and we bring the resulting equation of motion for
$ \Delta \bmath{x} $ to the form
$$
\left( { { {{\rm d}^2} \over {{\rm d}t^2} } + {n_0}^2 } \right)
\Delta \bmath{x}
= - { { \upartial V_1 } \over { \upartial \bmath{x} } } . \eqno\stepeq
$$
In the integration of equation (30) in what follows, we shall make use of
the requirement that $ \bmath{x} $ and $ \bmath{v} $ satisfy the
unperturbed equations of motion
$$
{ { {\rm d} \bmath{x} } \over {\rm d}t } = \bmath{v} \quad \hbox {and} \quad
{ { {\rm d} \bmath{v} } \over {\rm d}t } = - {n_0}^2 \bmath{x} . \eqno\stepeq
$$

For a radial oscillation of the system, the Eulerian perturbation
of the gravitational potential is of the form
$ V_1 ( \bmath{x} , t ) = V_1 (r) \exp ( - {\rm i} \omega t )$, where
$\omega$ is the frequency of the oscillation.  When the radial function
$V_1 (r)$ is imposed in the form of a polynomial in $r^2$, the solution of
equation (30) is a superposition of the solutions of a set of
equations of the form
$$
\left( { { {{\rm d}^2} \over {{\rm d}t^2} } + {n_0}^2 } \right)
\Delta \bmath{x}
= - \bmath{x} r^{2k} \exp ( - {\rm i} \omega t )
\quad (k= 0,1,2 \dots .) . \eqno\stepeq
$$

For the integration of equations (32), we introduce the variables
$$
{\bmath{z}}_{\sigma} = \bmath{x} + \sigma {\rm i} {n_0}^{-1} \bmath{v}
\quad ( \sigma = +1,-1) . \eqno\stepeq
$$
After writing the right-hand side of equation (32) in terms of the
new variables and expanding the resulting expression, we bring equation
(32) to the form
$$
\left( { { {{\rm d}^2} \over {{\rm d}t^2} } + {n_0}^2 } \right)
\Delta \bmath{x}
= - \left( 1 \over 4 \right)^k \sum_{n=0}^k \sum_{p=0}^{k-n} \sum_{\sigma}
2^{n-1} \left(\matrix{ k \cr n\cr }\right) 
\left(\matrix{ k-n \cr p \cr }\right) \bmath{z}_\sigma
M({\bmath{z}}_{+1} , {\bmath{z}}_{-1}; k-n-p,n,p) \exp (-{\rm i} \omega t ) ,
\eqno\stepeq
$$
where $({}_n^m )$ denotes a binomial coefficient,
the sum over $\sigma$ is over the values $ \sigma = +1 $ and
$ \sigma = -1 $,
$$
M({\bmath{z}}_{+1} , {\bmath{z}}_{-1}; m,n,p) = 
( {\bmath{z}}_{+1} \cdot {\bmath{z}}_{+1} )^m
( {\bmath{z}}_{+1} \cdot {\bmath{z}}_{-1} )^n
( {\bmath{z}}_{-1} \cdot {\bmath{z}}_{-1} )^p , \eqno\stepeq
$$
and $m (= k - n - p)$, $n$, and $p$ are integers.  Therefore, the solution of
equation (32) can be written as a superposition of solutions of
the equations 
$$
\left( { { {{\rm d}^2} \over {{\rm d}t^2} } + {n_0}^2 } \right)
\Delta \bmath{x}
= - {\bmath{z}}_{\sigma} M({\bmath{z}}_{+1} , {\bmath{z}}_{-1}; m,n,p)
\exp ( - {\rm i} \omega t ) .
\eqno\stepeq
$$

In virtue of equations (31), the variables ${\bmath{z}}_{\sigma}$
satisfy the equations of motion
$$
{ { {\rm d} {\bmath{z}}_{\sigma} } \over {\rm d}t } =
- \sigma {\rm i} n_0
{\bmath{z}}_{\sigma} \quad (\sigma = +1,-1) . \eqno\stepeq
$$
With the aid of equations (35) and (37), we readily verify that the
particular solution of equation (36) for the response of the system to
an imposed \lq{Eulerian perturbation of the acceleration}\rq\/ of the form
$  - {\bmath{z}}_{\sigma} M({\bmath{z}}_{+1} , {\bmath{z}}_{-1}; m,n,p)
\exp ( - {\rm i} \omega t ) $ is
$$
\Delta \bmath{x}_\sigma ({\bmath{z}}_{+1} , {\bmath{z}}_{-1}, t; m,n,p)
= { { {\bmath{z}}_{\sigma} 
M({\bmath{z}}_{+1} , {\bmath{z}}_{-1}; m,n,p) \exp ( - {\rm i} \omega t ) }
\over { [ \omega + ( 2m - 2p + \sigma ) n_0 ]^2 - {n_0}^2 } } .
\eqno\stepeq
$$
The Lagrangian perturbation of the velocity is accordingly
$$
\Delta \bmath{v}_\sigma ({\bmath{z}}_{+1} , {\bmath{z}}_{-1}, t; m,n,p)
 = { { {\rm d} } \over {\rm d}t } 
\Delta \bmath{x}_\sigma ({\bmath{z}}_{+1} , {\bmath{z}}_{-1}, t; m,n,p)
= -{\rm i}[ \omega + ( 2m - 2p + \sigma ) n_0 ]
\Delta \bmath{x}_\sigma ({\bmath{z}}_{+1} , {\bmath{z}}_{-1}, t; m,n,p).
\eqno\stepeq
$$
The Lagrangian displacement described by equations (38) and (39)
can now be represented as the six-vector
$$
\eqalign{
\boldeta_\sigma ({\bmath{z}}_{+1} , {\bmath{z}}_{-1}, t; m,n,p)
&= \Bigl(\matrix{
\Delta \bmath{x}_\sigma ({\bmath{z}}_{+1} , {\bmath{z}}_{-1}, t; m,n,p) \cr
\Delta \bmath{v}_\sigma ({\bmath{z}}_{+1} , {\bmath{z}}_{-1}, t; m,n,p)\cr }
\Bigr) \cr
&= { 1 \over { [ \omega + ( 2m - 2p + \sigma ) n_0 ]^2 - {n_0}^2 } }
\Bigl(\matrix { {\bmath{z}}_{\sigma}
M({\bmath{z}}_{+1} , {\bmath{z}}_{-1}; m,n,p)  \cr
         { -{\rm i}[ \omega + ( 2m - 2p + \sigma ) n_0 ] } {\bmath{z}}_{\sigma}
M({\bmath{z}}_{+1} , {\bmath{z}}_{-1}; m,n,p) \cr }
 \Bigr)
{  \exp ( - {\rm i} \omega t ) . } \cr } \eqno\stepeq
$$

For the purpose of writing the solutions of equations (32), it is
useful to rewrite equation (40) in the form
$$
\boldeta_\sigma ({\bmath{z}}_{+1} , {\bmath{z}}_{-1}, t; m,n,p) =
- { { \sigma
{ \bmu}_{\sigma}^{+1} ({\bmath{z}}_{+1} , {\bmath{z}}_{-1} ;m,n,p)
{  \exp ( - {\rm i} \omega t ) } }
\over { 2 n_0 [ \omega + (2m-2p+2 \sigma )n_0 ] } }
+ { { \sigma
{ \bmu}_{\sigma}^{-1} ({\bmath{z}}_{+1} , {\bmath{z}}_{-1} ;m,n,p)
{  \exp ( - {\rm i} \omega t ) } }
\over { 2 n_0 [ \omega + (2m-2p)n_0 ] } } ,
\eqno\stepeq
$$
where the quantities
$$
{ \bmu}_{\sigma}^{\tau} ({\bmath{z}}_{+1} , {\bmath{z}}_{-1} ;m,n,p) = 
\left(\matrix { {\bmath{z}}_{\sigma}
M({\bmath{z}}_{+1} , {\bmath{z}}_{-1}; m,n,p)  \cr
         { \tau \sigma {\rm i} n_0 } {\bmath{z}}_{\sigma}
M({\bmath{z}}_{+1} , {\bmath{z}}_{-1}; m,n,p) \cr }
 \right)   \quad
( \sigma = +1,-1; \tau = +1,-1)
\eqno\stepeq
$$
are \lq{six-vector monomials}\rq\/  in the three-vectors
$ \bmath{z}_\sigma $.  Equation (41) represents the required
solution of equation (36).  Upon comparing equations (34) and
(36), we see that the solution of equation (34) is
$$
\boldeta ({\bmath{z}}_{+1} , {\bmath{z}}_{-1}, t; k) =
 - \left( 1 \over 4 \right)^{k+1} \sum_{n=0}^k \sum_{p=0}^{k-n} 
\sum_{\sigma , \tau} 2^n \left(\matrix{ k \cr n\cr }\right) 
\left(\matrix{ k-n \cr p \cr }\right)
\left( { \tau \sigma 
\bmu_\sigma^\tau ({\bmath{z}}_{+1} , {\bmath{z}}_{-1} ;k-n-p,n,p)
\exp ( - {\rm i} \omega t) } \over
{ n_0 [ \omega + (2k-2n-4p + \sigma + \tau\sigma) n_0 ] }
\right) , \eqno\stepeq
$$
a superposition of the Lagrangian displacements represented
by equation (41), where the sum over $\tau$ is over the values
$\tau=+1$ and $\tau=-1$.  In other words, equation (43) represents
the Lagrangian displacement that satisfies equation (32) and
describes the response of a homogeneous sphere to an \lq{Eulerian
perturbation of the potential}\rq\/ equal to
${{(2k+2)}^{-1}}r^{2k+2} {\rm exp} ( - {\rm i} \omega t)$.

\subsection{Vector polynomials and their essential properties}

For $k = 0,1,\dots$, we identify the vector polynomials to be used in
the construction of the basis vectors by inspection of equation (43).
For a given value of $k$, we observe that the possible values of the
denominators on the right-hand side of equation (43) are 
$ \omega + (2k-2n-4p + \sigma + \tau\sigma) n_0 = \omega + (2k+2) n_0 $,
$ \omega + 2kn_0 $, $ \omega + (2k-2)n_0 , \dots $, $\omega - (2k+2)n_0 $,
apart from a common factor $ n_0 $.  The number of distinct values
of such denominators for a given value of $k$ is equal to $2k+3$.
Now we have obtained equation (43) as a solution of equation (30)
(see also equation [32]) for an arbitrarily imposed Eulerian
perturbation of the gravitational acceleration.  In particular,
the frequency $\omega$ has an arbitrary value.  On the other hand,
the basis vectors and their ingredients should be independent of the
value of $\omega$.  Therefore, we suppress the factor $\exp ( - i \omega t ) $
on the right-hand side of equation (43), and we identify the sum of
the numerators there for each distinct value of the denominator
$ \omega + (2k-2n-4p + \sigma + \tau\sigma) n_0 $ as a vector polynomial. 
There are thus $2k+3$ vector polynomials for a given power $k$ of $ r^2$
on the right-hand side of equation (32).  This procedure determines
the vector polynomials apart from arbitrary normalization factors.

In the case that $k = 0$, for example, equation (43) can be written
explicitly as
$$
\eqalign{
\boldeta ({\bmath{z}}_{+1} &, {\bmath{z}}_{-1}, t; 0) \cr
&= - { 1 \over {4 n_0 }}
\Bigl( { { \bmu_{+1}^{+1} ({\bmath{z}}_{+1} , {\bmath{z}}_{-1} ;0,0,0) }
\over { \omega + 2n_0 } }
- { { \bmu_{+1}^{-1} ({\bmath{z}}_{+1} , {\bmath{z}}_{-1} ;0,0,0) }
\over { \omega } }
- { { \bmu_{-1}^{+1} ({\bmath{z}}_{+1} , {\bmath{z}}_{-1} ;0,0,0) }
\over { \omega - 2n_0 } }
+ { { \bmu_{-1}^{-1} ({\bmath{z}}_{+1} , {\bmath{z}}_{-1} ;0,0,0) }
\over { \omega } } \Bigr) , \cr} \eqno\stepeq
$$
where we have suppressed the factor $\exp ( - {\rm i} \omega t)$ on the
right-hand side.
An inspection of the right-hand side of equation (44) as described
in the preceding paragraph yields the three vector polynomials
$$
\eqalign{ &\matrix{ \bpi({\bmath{z}}_{+1} , {\bmath{z}}_{-1} ;0,1)
= \bmu_{+1}^{+1} ({\bmath{z}}_{+1} , {\bmath{z}}_{-1} ;0,0,0) ,&
\bpi({\bmath{z}}_{+1} , {\bmath{z}}_{-1} ;0,2)
= \bmu_{-1}^{+1} ({\bmath{z}}_{+1} , {\bmath{z}}_{-1} ;0,0,0) ,\cr}
\quad \hbox{ and }\cr
&\bpi({\bmath{z}}_{+1} , {\bmath{z}}_{-1} ;0,3)
= \bmu_{-1}^{-1} ({\bmath{z}}_{+1} , {\bmath{z}}_{-1} ;0,0,0)
- \bmu_{+1}^{-1} ({\bmath{z}}_{+1} , {\bmath{z}}_{-1} ;0,0,0) . \cr}
\eqno\stepeq
$$
In the notation $ \bpi({\bmath{z}}_{+1} , {\bmath{z}}_{-1} ;k,h) $
for a vector polynomial, we adopt the convention that the integer $k$
identifies the {\it group} of vector polynomials to which
$ \bpi({\bmath{z}}_{+1} , {\bmath{z}}_{-1} ;k,h) $ belongs,
i.e., the power $k$ of $r^2$ on the right-hand side of equation (32), and
the value of the integer $h$ $(h = 1,2,\dots,2k+3)$ distinguishes
$ \bpi({\bmath{z}}_{+1} , {\bmath{z}}_{-1} ;k,h) $ from the other members
of that group.  In equations (A1)-(A3) in Appendix A, we list the
expressions for the additional vector polynomials through the group $k=3$.
We choose the normalization factors in the vector
polynomials listed in equations (45) and (A1)-(A3) and we assign
the order in which to list the vector polynomials in each group
so that most of the vector polynomials are conveniently arranged in
complex-conjugate pairs (see eq. [51] below).

The vector polynomials listed in equations (45) and (A1)-(A3) would be
ingredients for an exact construction of normal modes in the homogeneous spheres
described by Fridman and Polyachenko (1984, see Chapter III, Section 4.1).
In order to make use of the six-vector monomials $ \bmu_\sigma^\tau (m,n,p) $
and the six-vector polynomials $ \bpi(k,h) $ in the representation
of perturbations in inhomogeneous
spheres, we must generalize the definition of $ n_0 $.  A suitable
generalization is provided by the equation
$$
{n_0}^2 = - { W \over I } , \eqno\stepeq
$$
where
$$
W = - \int_V \rho_0 (r) r { { \upartial V_0 (r) } \over { \upartial r } }
{\rm d} \bmath{x} \quad \hbox { and } I = \int_V \rho_0 (r) r^2
{\rm d} \bmath{x} 
\eqno\stepeq
$$
denote the gravitational potential energy and the moment of inertia,
respectively, of the unperturbed configuration, $ \rho_0 (r)$ and
$V_0 (r)$ are the density and gravitational potential, respectively,
of the unperturbed configuration, and the integrations extend over
the volume $V$ of the unperturbed configuration.  For homogeneous
spheres, equations (28) and (46) are equivalent definitions of $n_0$.
In general, however, the variables $ \bmath{z}_\sigma $,
the scalar monomials $ M(\bmath{z}_{+1},\bmath{z}_{-1};m,n,p) $,
the vector monomials 
$ \bmu_\sigma^\tau (\bmath{z}_{+1},\bmath{z}_{-1};m,n,p) $, and the
vector polynomials $ \bpi (\bmath{z}_{+1},\bmath{z}_{-1};k,h) $
$(k=0,\dots,3)$ are defined by equations (33), (35), (42), and (45) and
(A1)-(A3), respectively, with $n_0$ defined by equation (46).

Making use of equations (33), (35), and (42) in explicit calculations
applied to the expressions for the vector polynomials listed in
equations (45) and (A1)-(A3), we find that
$$
\bnabla_6 \cdot \bpi (k,h) = 0 . \eqno\stepeq
$$
Thus, the construction of the basis vectors from the vector polynomials
will ensure that the representation of the Lagrangian displacement in
equation (25) will automatically satisfy equation (10). 

In the calculations that follow, we shall need the complex conjugates
of the vector polynomials and the vectors adjoint to the vector
polynomials.  These are to be constructed in terms of the vector
monomials, their complex conjugates and their adjoints.  Referring
to equations (33), (35), and (42), respectively, we find that
$$
\bmath{z}_\sigma^* = \bmath{z}_{- \sigma}, \quad
M^* (\bmath{z}_{+1},\bmath{z}_{-1};m,n,p) = 
M(\bmath{z}_{+1},\bmath{z}_{-1};p,n,m) , \quad  \hbox {and} \quad
[ \bmu_\sigma^\tau (\bmath{z}_{+1},\bmath{z}_{-1};m,n,p) ]^* =
\bmu_{-\sigma}^\tau (\bmath{z}_{+1},\bmath{z}_{-1};p,n,m) ,
\eqno\stepeq
$$
where the asterisk, written as a superscript, labels the complex
conjugate of a quantity.  With the aid of equations (16)-(18),
(33), and (49), we can also verify that
$C [ \bmath{z}_\sigma M({\bmath{z}}_{+1} , {\bmath{z}}_{-1}; m,n,p) ]
 = \bmath{z}_\sigma M({\bmath{z}}_{+1} , {\bmath{z}}_{-1}; m,n,p) $
and, accordingly, that
$$
A{ \bmu}_{\sigma}^{\tau} ({\bmath{z}}_{+1} , {\bmath{z}}_{-1} ;m,n,p) =
{ \left(\matrix{0&C\cr
           C&0\cr}\right) }
{ \left(\matrix { {\bmath{z}}_{\sigma}
M({\bmath{z}}_{+1} , {\bmath{z}}_{-1}; m,n,p)  \cr
         { \tau \sigma {\rm i} n_0 } {\bmath{z}}_{\sigma}
M({\bmath{z}}_{+1} , {\bmath{z}}_{-1}; m,n,p) \cr } \right) }
 = {  \left(\matrix { - { \tau \sigma {\rm i} n_0 } {\bmath{z}}_{\sigma}
M({\bmath{z}}_{+1} , {\bmath{z}}_{-1}; m,n,p)  \cr
{\bmath{z}}_{\sigma} M({\bmath{z}}_{+1} , {\bmath{z}}_{-1}; m,n,p) \cr }
 \right) } .
\eqno\stepeq
$$

The third of equations (49) implies that the vector polynomials listed
in equations (45) and (A1)-(A3) satisfy the complex-conjugate relations
$$
\bpi(k,h) = \bpi^* (k,h-1) \quad (h=2,4,\dots,2k+2) \quad \hbox {and}
\quad \bpi (k,2k+3) = - \bpi^* (k,2k+3) . \eqno\stepeq
$$
In other words, the group of vector polynomials for a given value of $k$
contains $k+1$ complex-conjugate pairs of members and one imaginary
member.

For the explicit construction of the basis vectors and for the associated
numerical calculations, we relabel the vector polynomials in
terms of a single integer.  The prescription for doing so depends on
the number of groups of vector polynomials to in used in the construction
of the basis vectors.  Thus, in the following subsection, we shall
construct the basis vectors from the groups of vector polynomials
$\bpi (k,h) $ for which $k = 0,1,\dots,k_{\rm max} $, say.

We first relabel the vector polynomials which form complex-conjugate
pairs by writing (see eqs. [51])
$$
\bpi(\bmath{z}_{+1} , \bmath{z}_{-1} ;k,h)
= \bpi(\bmath{z}_{+1} , \bmath{z}_{-1};q) , \quad \hbox {where} \quad
q=k(k+1) + h \quad (k=0,1,\dots,k_{\rm max};h=1,2,\dots,2k+2) . \eqno\stepeq
$$
The value of the new integer $q$ to be associated with a given pair of
values $(k,h)$ is determined here as follows.  We recall that the
group of vector polynomials $\bpi(k^\prime , h^\prime )$, for a given value of
$k^\prime$ $(k^\prime < k)$, contains $2k^\prime +2$ complex members.
Summing that number from $k^\prime = 0$ to $k^\prime = k-1$, we count a
total of $k(k+1)$ complex vector polynomials with $0 \le k^\prime \le k-1$.
We then let $q = k(k+1) + h$, inasmuch as the polynomial $\bpi(k,h)$
is number $h$ in the listing of the polynomials belonging to the group $k$.
The integer $q$ takes the values $ q = 1,2,\dots,Q$ in equations (52),
where $Q = (k_{\rm max} + 1)(k_{\rm max} + 2) $ is the total number of
complex vector polynomials in the groups such that $0 \le k \le k_{\rm max}$.
We next add the purely imaginary vector polynomials to the enumeration
in equation (52) by writing
$$
\bpi(\bmath{z}_{+1} , \bmath{z}_{-1} ;k,2k+3)
= \bpi(\bmath{z}_{+1} , \bmath{z}_{-1};q) , \quad \hbox {where} \quad
q= Q + k + 1 \qquad  (k=0,1,\dots,k_{\rm max}) . \eqno\stepeq
$$
Here, the expression for $q$ is determined by the circumstances that we are
continuing the enumeration of vector polynomials begun in equation
(52) and each group of vector polynomials contributes one purely imaginary member
to the enumeration (see eq.[51]). The values of the integer $q$ in equation (53) are
$q=Q+1,Q+2,\dots,N$, where
$N=Q+k_{\rm max} + 1 = (k_{\rm max} + 1)(k_{\rm max} + 3)$ is the
total number of vector polynomials, complex and purely imaginary, in
the groups such that $0 \le k \le k_{\rm max}$.

Rewriting equations (51) in accordance with equations (52) and (53),
we bring the complex-conjugate relations satisfied by the vector
polynomials to the form
$$
\bpi (q) = \bpi^* (q-1) \quad
(q=2,4,\dots,Q) \quad \hbox {and} \quad
\bpi(q) = - \bpi^* (q) \qquad 
(q=Q +1,Q +2, \dots,N) . \eqno\stepeq
$$

According to equations (45) and (A1)-A3), the vector polynomials are
superpositions of vector monomials of the form
$$
\bpi(\bmath{z}_{+1} , \bmath{z}_{-1} ;q) = \sum_{j=1}^J \langle q|j \rangle 
\bmu_\sigma^\tau (\bmath{z}_{+1} , \bmath{z}_{-1} ;m,n,p) , \eqno\stepeq
$$
where we label the vector polynomials
$\bpi(\bmath{z}_{+1} , \bmath{z}_{-1} ;q)$ in terms of the single integer
$q$ in accordance with equations (52) and (53).  On the right-hand
sides of equations (55), we determine the constant coefficients
$\langle q|j \rangle$, we determine the integers which distinguish the
vector monomials 
$\bmu_\sigma^\tau (\bmath{z}_{+1} , \bmath{z}_{-1} ;m,n,p)$ as functions
$\sigma = \sigma (q;j)$, $\tau = \tau (q;j)$, $m=m (q;j)$, $n=n (q;j)$,
and $p=p (q;j)$, and we determine the number
of terms $J=J (q)$ in each sum by inspection of equations (45) and
(A1)-A3).  The coefficients $\langle q|j \rangle$ are seen there to
be real.  It follows that
$$
\bpi^* (\bmath{z}_{+1} , \bmath{z}_{-1} ;q) = \sum_{j=1}^J \langle q|j \rangle 
[\bmu_\sigma^\tau (\bmath{z}_{+1} , \bmath{z}_{-1} ;m,n,p)]^* 
\quad  \hbox {and} \quad
A\bpi(\bmath{z}_{+1} , \bmath{z}_{-1} ;q) = \sum_{j=1}^J \langle q|j \rangle 
A\bmu_\sigma^\tau (\bmath{z}_{+1} , \bmath{z}_{-1} ;m,n,p) . \eqno\stepeq
$$

\subsection{Construction of the basis vectors}

We now write the basis vectors in the form
$$
\bbeta (1) = N(1) \bpi(1) \quad \hbox {and} \quad \bbeta  (q) =
N(q) \left[ \sum_{r=1}^{q-1} c(q,r) \bbeta (r) + \bpi(q) \right]
\qquad (q=2,3,\dots,N) , \eqno\stepeq
$$
where the constants $N(q)$ and $c(q,r)$ are to be determined with the
aid of a Gram-Schmidt process.  In that process,  we require that the basis
vectors be orthogonal in the sense of equation (24).  Thus, for example,
when we form the inner products of $A \bbeta (s)$ $(s \le q)$ with both
sides of the second of equations (57) and apply the orthogonality
conditions, we obtain
$$
\eqalign{
N(q) \left[ { \langle A \bbeta (s), \bbeta (s) \rangle }_2 c(q,s)
+ { \langle A \bbeta (s), \bpi (q) \rangle }_2 \right] = 0
\quad &\hbox {and} \quad
{ \langle A \bbeta (q), \bbeta (q) \rangle }_2 = 
N(q) { \langle A \bbeta (q), \bpi (q) \rangle }_2 \cr
(q=2,3,\dots,N;&s=1,2,\dots,q-1) . \cr }
\eqno\stepeq
$$

For the reduction of equations (58) to forms suitable for the systematic
evaluation of $N(q)$ and $c(q,s)$, we need the equations
$$
A \bbeta (1) = N^* (1) A \bpi(1) \quad \hbox {and} \quad A \bbeta  (q) =
N^* (q) \left[ \sum_{r=1}^{q-1} c^* (q,r) A \bbeta (r) + A \bpi(q) \right]
\qquad (q=2,3,\dots,N) \eqno\stepeq
$$
for the vectors adjoint to the basis vectors $\bbeta (q)$.  We obtain
equations (59) by transforming equations (57) with the aid of
equations (17) and (18).  Equations (57) describe a representation
of the basis vectors in which the constants $N(q)$ must not vanish.
Consequently the second of equations (59) allows us to reduce the first
of equations (58) (except when $q=2$ and $s=1$) to
$$
{ \langle A \bbeta (s), \bbeta (s) \rangle }_2 c(q,s)
= - N(s) \left[ \sum_{r=1}^{s-1} c(s,r)
{ \langle A \bbeta (r), \bpi (q) \rangle }_2
+ { \langle A \bpi (s), \bpi (q) \rangle }_2 \right] \qquad
(q=3,4,\dots,N;s=2,3,\dots,q-1) .
\eqno\stepeq
$$
The first of equations (58), again with the common factor $N(q)$
suppressed, enables us to rewrite equation (60) as
$$
{ \langle A \bbeta (s), \bbeta (s) \rangle }_2 c(q,s)
= N(s) \left[ \sum_{r=1}^{s-1}
{ \langle A \bbeta (r), \bbeta (r) \rangle }_2 c(s,r)c(q,r) 
- { \langle A \bpi (s), \bpi (q) \rangle }_2 \right] \qquad
(q=3,4,\dots,N;s=2,3,\dots,q-1) . \eqno\stepeq
$$
Likewise, with the aid of the second of equations (59) and then the
first of equations (58), we reduce the second of equations (58) to the form
$$
{ \langle A \bbeta (q), \bbeta (q) \rangle }_2
= N^2 (q) \left[ { \langle A \bpi (q), \bpi (q) \rangle }_2
- \sum_{r=1}^{q-1}  { \langle A \bbeta (r), \bbeta (r) \rangle }_2
c^2 (q,r) \right] \qquad (q=2,3,\dots,N) . \eqno\stepeq
$$
Finally, making use of the first of equations (59) in order to
rewrite the first of equations (58) in the case that $s=1$ and then
the second of equations (58) in the case that $q=1$, we obtain
$$
{ \langle A \bbeta (1), \bbeta (1) \rangle }_2 c(q,1)
= - N(1) { \langle A \bpi (1), \bpi (q) \rangle }_2
\qquad (q=2,3,\dots,N) \quad \hbox { and } \quad
{ \langle A \bbeta (1), \bbeta (1) \rangle }_2
= N^2 (1) { \langle A \bpi (1), \bpi (1) \rangle }_2
\eqno\stepeq
$$
as special cases of equations (61) and (62), respectively.  The inner
products $ { \langle A \bpi (q^\prime ), \bpi (q) \rangle }_2 $ in
equations (61)-(63) are given quantities determined by the unperturbed
structure of the system (see eq.[9]).  Expressions required for the
evaluation of $ { \langle A \bpi (q^\prime ), \bpi (q) \rangle }_2 $
are derived in Appendix B, and the symmetries and other important
properties of $ { \langle A \bpi (q^\prime ), \bpi (q) \rangle }_2 $
are described there.

Equations (61)-(63) are the basic equations for the determination of the
constants $N(q)$ and $c(q,s)$ in the representation of the basis vectors
introduced in equations (57).  Before we can solve those equations, we
must normalize the basis vectors.  In what follows, we shall find that
the appropriate normalization is
$$
\eqalign{
{ \langle A \bbeta (q), \bbeta (q) \rangle }_2 &= 1
\qquad (q = 1,2,\dots,Q) \cr
&= 0 \qquad ( q = Q +1, Q + 2, \dots,N ) . \cr} \eqno\stepeq
$$
The unusual result that some of the inner products
${ \langle A \bbeta (q), \bbeta (q) \rangle }_2$ must vanish
complicates the matrix method somewhat, but it seems not to
create fundamental difficulties in the application of the matrix
method.

With the normalization of the basis vectors given, we can solve
equations (61)-(63) for the quantities $ N(q)$ and $c(q,s)$
in the following sequence.  We first evaluate $N(1)$ and $c(q,1)$ 
$(q=2,3,\dots,N)$ in accordance with
equations (63), and we next evaluate $N(2)$ in accordance with the first of
equations (62).  We subsequently solve equations (61) and (62) in
the sequence $q=3,4,\dots,N$.  For each
value of $q$, in turn, we evaluate the constants $c(q,s)$ in the sequence
$s=2,3,\dots,q-1$ with the aid of equations (61), and we finally
evaluate $N(q)$ in accordance with equation (62).  At each step in this
process, the values of $c(s,r)$ and $N(s)$ required for the evaluation
of the right-hand side of equation (61) or the coefficient of $N^2 (q)$
on the right-hand side of equation (62) are known as results
of earlier steps in the process.

In Sections 4.3.1 and 4.3.2 which follow, we present separate
discussions of the solutions of equations (61)-(63) for the quantities
$N(q)$ and $c(q,s)$ involved in the representations of the basis vectors
satisfying the different normalization conditions in equations (64).
The solutions for $N(q)$ and $c(q,s)$ satisfy certain identities
which are derived in Appendix C.  Some of the results
described in Sections 4.3.1 and 4.3.2 and in Appendix C follow,
because the solutions of equations (61)-(63) have the property
that the quantities $N^2 (q)$, $c(q,s)/N(s)$, $c(s,r)c(q,r)$, and
$c^2 (q,r)$ are all imaginary.  For a proof of this result, we note
that the inner products ${ \langle A \bpi (s), \bpi (q) \rangle }_2$
are imaginary quantities, as is shown in Appendix B, and we observe
that we have normalized the basis vectors so that the inner products
${ \langle A \bbeta (s), \bbeta (s) \rangle }_2$ are real quantities.
Noting that equations (63) establish the results claimed here for
$N(1)$ and $c(q,1)$, we can prove the general result by induction with
the aid of equations (61) and (62).

\subsubsection{Complex-conjugate pairs of basis vectors}

When $0<q \le Q $ and $0<q^\prime \le Q $, the inner products
$ { \langle A \bpi (q^\prime ), \bpi (q) \rangle }_2 $ do
not vanish, in general.  With the normalization given in
equations (64), we find that we can satisfy equations (61)-(63)
consistently when $0<q \le Q$.  In particular, we can solve equations (63)
in the manner
$$
 N^2 (1) = \left[ { \langle A \bpi (1), \bpi (1) \rangle }_2 \right]^{-1}
\quad \hbox {and} \quad
c(q,1) = - N(1) { \langle A \bpi (1), \bpi (q) \rangle }_2
\qquad (q=2,3,\dots,Q) .
\eqno\stepeq
$$
We can also bring the subsets of equations (61) and (62)
with $q \le Q $ to the forms
$$
c(q,s)
= N(s) \left[ \sum_{r=1}^{s-1} c(s,r)c(q,r) 
- { \langle A \bpi (s), \bpi (q) \rangle }_2 \right]
\qquad (q = 3,4,\dots,Q ;
s=2,3,\dots,q-1) \eqno\stepeq
$$
and
$$
N^2 (q) = { \left[ { \langle A \bpi (q), \bpi (q) \rangle }_2
- \sum_{r=1}^{q-1}  c^2 (q,r) \right] }^{-1}
\qquad (q=2,3,\dots,Q) , \eqno\stepeq
$$
respectively.  Equations (65)-(67) enable us to evaluate the constants
$c(q,s)$ and $N(q)$, in the case that $q \le Q$,
in the sequence prescribed above in the paragraph following the
normalization of the basis vectors in equations (64).

An important consequence of equations (65)-(67) is that the
basis vectors $\bbeta (q)$ $(q \le Q)$ form complex-conjugate pairs in the
arrangement
$$
\bbeta (q) = \bbeta^* (q-1) \qquad (q=2,4,\dots,Q) . \eqno\stepeq
$$
The expressions for the basis vectors given by equations (57)
satisfy equations (68) because the vector polynomials $\bpi (q)$
satisfy equations (54) and the quantities determined by equations
(65)-(67) satisfy the identities
$$
\eqalign{
&N(q) = N^* (q-1) \quad \hbox {and} \quad c(q,q-1)=0
\qquad (q=2,4,\dots,Q)\quad \hbox {and} \cr 
&c(q,s) = c^* (q-1,s-1) \quad \hbox {and} \quad
c(q,s-1) = c^* (q-1,s) \qquad
(q=4,6,\dots,Q;s=2,4,\dots,q-2) . \cr }
\eqno\stepeq
$$
These identities are verified in Appendix C.

\subsubsection{Null basis vectors}

It remains to solve equations (61)-(63) in the case that
$Q < q \le N$.  As an essential part of that
solution, we must show that the resulting basis vectors are
null vectors as we have claimed in equation (64).

When $s \le Q < q \le N$, the normalization of the
basis vectors ${ \langle A \bbeta (s), \bbeta (s) \rangle }_2 = 1$
$(s = 1,2,\dots,Q)$ reduces the first of equations (63) and equations
(61) to
$$
c(q,1) = - N(1) { \langle A \bpi (1), \bpi (q) \rangle }_2
\qquad (q=Q+1,Q+2,\dots,N)
\eqno\stepeq
$$
and
$$
c(q,s)
= N(s) \left[ \sum_{r=1}^{s-1} c(s,r)c(q,r) 
- { \langle A \bpi (s), \bpi (q) \rangle }_2 \right]
\qquad (q=Q+1,Q+2,\dots,N ; s=2,3,\dots,Q) , \eqno\stepeq
$$
respectively.  Starting with the known values of $c(q,r)$ and $N(q)$,
evaluated, as described Section 4.3.1, with the aid of
equations (65)-(67), we can
make use of equations (70) and (71) in order to evaluate the additional
quantities $c(q,s)$ in a sequence such that, at each step in the process,
the values of quantities required for the evaluation of the right-hand side
of equation (71) are known as results of preceding steps.

We must finally consider the solution of equations (61) and (62) in the
case that $Q<s<q \le N$.  In this case, the inner products
${ \langle A \bpi (s), \bpi (q) \rangle }_2$ and
${ \langle A \bpi (q), \bpi (q) \rangle }_2$ vanish according to equations (B7).
The further reduction of equations (61) and (62) requires additional identities,
especially equations (C11), which are also derived in Appendix C.
In virtue of the normalization
${ \langle A \bbeta (q), \bbeta (q) \rangle }_2 = 1$
$(q = 1,2,\dots,Q)$ and equations (C11), the sums of terms from
$r=1$ to $r=Q$ vanish on the right-hand
sides of of equations (61) and (62).  Thus, the subset of
equations (61) and (62) which we must finally consider reduces to
$$
{ \langle A \bbeta (s), \bbeta (s) \rangle }_2 c(q,s)
= N(s) \sum_{r=Q+1}^{s-1}
{ \langle A \bbeta (r), \bbeta (r) \rangle }_2 c(s,r)c(q,r)
\qquad (q=Q+2,Q+3,\dots,N;s=Q+1,Q+2,\dots,q-1) , \eqno\stepeq
$$
$$
{ \langle A \bbeta (Q+1), \bbeta (Q+1) \rangle }_2 = 0 ,
\eqno\stepeq
$$
and
$$
{ \langle A \bbeta (q), \bbeta (q) \rangle }_2
= - N^2 (q) \sum_{r=Q+1}^{q-1}  { \langle A \bbeta (r), \bbeta (r) \rangle }_2
c^2 (q,r) \qquad (q=Q+2,Q+3,\dots,N) . \eqno\stepeq
$$

According to equation (74), if
${ \langle A \bbeta (r), \bbeta (r) \rangle }_2 = 0$ $(r=Q+1,\dots ,
q-1)$, then ${ \langle A \bbeta (q), \bbeta (q) \rangle }_2 = 0$.
Equation (73) shows that the required condition is satisfied for $q=Q+1$.
Therefore, by induction, we have ${ \langle A \bbeta (q), \bbeta (q) \rangle }_2 = 0$
$(q=Q+1,Q+2,,\dots,N)$ as is claimed in equation (64).
Consequently, equations (72) and (74) leave the values of the
remaining quantities $c(q,s)$ and $N(q)$ indeterminate.  Without
loss of generality, we assign those values in the manner
$$
c(q,s) = 0 \qquad (q=Q+2,Q+3,\dots,N;s=Q+1,Q+2,\dots,q-1)
\eqno\stepeq
$$
and
$$
N(q) = 1 + {\rm i} \qquad (q=Q+1,Q+2,\dots,N) . \eqno\stepeq
$$
The values of $N(q)$ adopted in equations (76) preserve the property
that the quantities $N^2 (q)$ are imaginary.

\subsection{The completeness of the basis vectors}

The question arises as to whether or not the set of basis vectors
constructed here is complete.  We have not established that basis
vectors constructed in this way form a complete set in general.
We can only say that the adopted set of basis vectors is
\lq sufficiently complete\rq\/ for the exact representation
of radial modes in certain homogeneous spheres.  Accordingly,
one might hope that they would suffice for the representation
of corresponding modes in inhomogeneous spheres.  It appears in
Section 6 that they do indeed suffice for the representation of
the lowest radial modes in rather centrally concentrated spheres.

The validity of the present formulation of the matrix method as a
method for the approximate solution of the characteristic value
problem governing modes does not require that the adopted basis
vectors form a complete set.  For the matrix method is a particular
realization of a variational method in which the basis vectors are
the ingredients of a trial function.  The essential condition to
be satisfied by the basis vectors is that they provide a sufficiently
accurate representation of the Lagrangian displacements of the modes
to be investigated.  This is essentially the point of view adopted
previously by Saha (1991).

\subsection{Construction of the matrix equations}

The construction of a suitable set of basis vectors reduces a given
application of the matrix method to the evaluation of the
matrix elements in a truncated set of equations (26)
and then to the solution of the matrix equations.  We describe the
construction of the required matrix elements in Appendix D.

\section{Solution of the characteristic value
problem in the matrix representation}

In the remainder of this paper, we shall restrict ourselves to the
consideration of radial oscillations in systems in which the distribution
function is a function of the energy alone and the velocity distribution
is isotropic.  The quadratures required for the evaluation of inner
products and matrix elements are described in Appendix E.
In what follows, we describe the reduction and solution of the matrix
equations.

\subsection{Reduction of the matrix equations}

With the truncation of the set of basis vectors to a finite set
containing $N$ members, and in virtue of the orthogonality and normalization
of the basis vectors specified in equations (24) and (64), respectively,
the matrix formulation of the characteristic value problem represented
in equation (26) now reduces to
$$
\omega a(k) 
=  \sum_{j=1}^N R^{(N)} (k,j)
 a(j) \qquad (k=1,2,\dots ,Q) \quad \hbox {and} \quad
0 =  \sum_{j=1}^N R^{(N)} (k,j)
 a(j) \qquad (k=Q+1,Q+2,\dots,N) ,  \eqno\stepeq
$$
where
$$
R^{(N)} (k,j) = {1 \over 2} 
[{ \langle { A { \bbeta } ( j ),P { \bbeta } ( k ) } \rangle}_2
 + { \langle { A { \bbeta } ( k ),P { \bbeta } ( j ) } \rangle}_2 ] .
\eqno\stepeq
$$
We reduce the system of equations (77) to a standard form by making
use of those members of the system in which $k=Q+1,Q+2,\dots,N$ in order
to eliminate the amplitudes $a(k)$ $(k=Q+1,Q+2,\dots,N)$ from
the remaining equations in the system.  We accomplish this with
the aid of a sequence
of transformations in each of which we eliminate one amplitude.  After
$N - K$ steps in that process, where $K$ $(Q+1 \le K \le N)$ is an integer,
we will have reduced equations (77) to a system of the form
$$
\omega a(k) 
=  \sum_{j=1}^K R^{(K)} (k,j)
 a(j) \qquad (k=1,2,\dots ,Q) \quad \hbox {and} \quad
0 =  \sum_{j=1}^K R^{(K)} (k,j)
 a(j) \qquad (k=Q+1,Q+2,\dots,K) ,  \eqno\stepeq
$$
say, where the matrix $R^{(K)}$ is derived from the matrix $R^{(N)}$
in a sequence of transformations of the form of equation (82) below.
Equation (82) is the transformation with which we eliminate $a(K)$.
Thus, from the last of equations (79), we derive
$$
a(K) = - {1 \over {R^{(K)} (K,K) }}
\sum_{j=1}^{K-1} R^{(K)} (K,j) a(j) .  \eqno\stepeq
$$
Upon substituting from this solution for $a(K)$ in equations (79)
$(k=1,2,\dots,K-1)$, we obtain the system
$$
\omega a(k) 
=  \sum_{j=1}^{K-1} R^{(K-1)} (k,j)
 a(j) \qquad (k=1,2,\dots ,Q) \quad \hbox {and} \quad
0 =  \sum_{j=1}^{K-1} R^{(K-1)} (k,j)
 a(j) \qquad (k=Q+1,Q+2,\dots,K-1) , \eqno\stepeq
$$
where
$$
R^{(K-1)} (k,j) = R^{(K)} (k,j)
- {{ R^{(K)} (k,K) R^{(K)} (K,j) } \over {R^{(K)} (K,K) }}.
\eqno\stepeq
$$
At the beginning of this sequence of transformations we have $K = N$,
and equations (79) coincide with equations (77).  At the end of the
sequence, we have $K = Q + 1$, and equations (81) reduce to
$$
\omega a(k) 
=  \sum_{j=1}^{Q} R^{(Q)} (k,j)
 a(j) \qquad (k=1,2,\dots ,Q) . \eqno\stepeq
$$
We are thus left with the characteristic value problem for the
$Q \times Q$ matrix $R^{(Q)} (k,j)$, where
$Q = (k_{\rm max} + 1)(k_{\rm max} + 2) $.

By definition (see eq. [78]) the matrix $R^{(N)} (k,j)$ has
the symmetry $R^{(N)} (j,k) = R^{(N)} (k,j)$.  It follows that
the matrices defined by equations (82) have the symmetry
$R^{(K)} (j,k) = R^{(K)} (k,j)$.  In particular,
$R^{(Q)} (j,k) = R^{(Q)} (k,j)$.  Moreover, in the numerical
investigation of the solutions of equation (83), we find that the matrix
$R^{(Q)} (k,j)$ has the following additional properties to the accuracy
in which that matrix is evaluated.

\beginlist
\item (i) For even values of $k - j$, the elements $R^{(Q)} (k,j)$ 
are real.
\item (ii) For odd values of $k - j$, the elements $R^{(Q)} (k,j)$ 
are imaginary.
\endlist

Thus the complex matrix $R^{(Q)} (k,j)$ can be expressed
in terms of a real matrix $S^{(Q)} (k,j)$, say, in the manner
$$
R^{(Q)} (k,j) = {\rm i}^{(k-j)} S^{(Q)} (k,j) . \eqno\stepeq
$$
If follows from this definition of $S^{(Q)} (k,j)$ and the symmetry
of $R^{(Q)} (k,j)$ that
$$
S^{(Q)} (k,j) = (-{\rm i})^{(k-j)} R^{(Q)} (k,j) \quad \hbox {and} \quad
S^{(Q)} (j,k) = (-1)^{(k-j)} S^{(Q)} (k,j) . \eqno\stepeq
$$
Letting
$$
a(k) = {\rm i}^k \alpha (k) \qquad (k=1,2,\dots,Q) , \eqno\stepeq
$$
where the quantities $ \alpha (k)$ are constants, we can now transform
the system of equations (83) to the system
$$
\omega \alpha (k) 
=  \sum_{j=1}^{Q} S^{(Q)} (k,j)
 \alpha (j) \qquad (k=1,2,\dots ,Q) , \eqno\stepeq
$$
and we are left with a characteristic value problem for the real
matrix $ S^{(Q)} (k,j) $.

The numerical calculations required for the evaluation of the matrix
$S^{(Q)} (k,j)$ and the solution of equations (87) have been performed
with the aid of programs written in FORTRAN 77 and run in single-precision
and complex arithmetic on Macintosh II and Macintosh IIcx computers.
The program for the solution of equations (87) incorporates subroutines
from the NAPAK library.  Numerical quadratures required in the evaluation
of inner products and matrix elements (see Appendix E) were performed with
the aid of the fourth order Runge-Kutta integrator described by
Press et al. (1986).

These calculations have been performed in the four approximations defined
by setting $k_{\rm max}$ equal to 0,1,2, and 3.  The corresponding values
of $Q = (k_{\rm max} + 1)(k_{\rm max} + 2)$ are 2, 6, 12, and 20,
respectively.  In other words, we have solved equations (87) for
2 by 2, 6 by 6, 12 by 12, and 20 by 20 matrices $S^{(Q)} (k,j)$,
respectively.  For an assigned value of $k_{\rm max}$, the number
of basis vectors included in the representation of the Lagrangian
displacement is $N = (k_{\rm max} + 1)(k_{\rm max} + 3)$.  Thus, we
include 3, 8, 15, and 24 basis vectors in the four approximations
considered here.  The numerical calculations required for the
evaluation of $S^{(Q)} (k,j)$ and the solution of equations (87) in
all four approximations require about four hours for one unperturbed
model.

\subsection{Characteristic vectors}

Let $\omega (n)$ $(n = 1,2,\dots,Q)$ denote the solutions of equations (87)
for the characteristic frequencies for a given value of $Q$, and let
$\alpha (j,n)$  $(j=1,2,\dots,Q;n=1,2,\dots,Q)$ denote the corresponding
solutions for the constants $\alpha (j)$.  The
solution for the Lagrangian displacement which belongs to the
characteristic frequency $\omega (n)$ is then
$$
{ \boldeta } (n) = \sum_{j=1}^N a(j,n) { \bbeta } ( j ) 
\eqno\stepeq
$$
(cf. eq. [25]).  Here,
$$
a(j,n) = {\rm i}^j \alpha (j,n) \qquad (j=1,2,\dots,Q)  \eqno\stepeq
$$
(cf. eq. [86]), and the remaining values of the constants $a(j,n)$
$(j=Q+1,Q+2,\dots,N)$ are determined by working backward
(i.e., in the sequence $K=Q+1,Q+2,\dots,N$) through equations
(80)-(82) with the aid of the values of the first $Q$
constants $a(j,n)$.

Equation (88) represents a characteristic
solution for the Lagrangian displacement in the
approximation that the right-hand side of equation (25) is
truncated to a superposition of $N$ basis vectors.  These characteristic
solutions satisfy orthogonality relations of the form
$$
{ \langle { A { \boldeta } ( m ), { \boldeta } ( n ) } \rangle}_2
= \delta_{mn} ,  \eqno\stepeq
$$
where $\delta_{mn}$ is the Kronecker delta.  In order to verify
equations (90), we first substitute from equations (88) for the
Lagrangian displacements in the inner product on the left-hand side
of equation (90) and bring the resulting expression to the form
$$
{ \langle { A { \boldeta } ( m ), { \boldeta } ( n ) } \rangle}_2
= \sum_{j=1}^Q a(j,m) a(j,n)  \eqno\stepeq
$$
with the aid of equations (24) and (64).  Now the requirement that
the quantities $\omega (n)$ and $\alpha (j,n)$ 
$(j=1,2,\dots,Q;n=1,2,\dots,Q)$ be a characteristic solution
of equations (87) implies that the quantities $\omega (n)$
and $a (j,n)$  $(j=1,2,\dots,Q;n=1,2,\dots,Q)$ must be
a characteristic solution of equations (83).  In the latter
system of equations, the matrix $R^Q (k,j)$ is symmetric in the
indices $k$ and $j$.  It therefore follows from equations (83)
that the constants $a(j,n)$ satisfy the conditions
$$
\sum_{j=1}^Q a(j,m) a(j,n) = \delta_{mn}.   \eqno\stepeq
$$
Without loss of generality, we have imposed here the normalization
condition that, when $m=n$, the sum on the left-hand side of
equation (92) is equal to unity.  Equations (91) and (92)
combine to establish equations (90).  These are precisely of
the form of the orthogonality relations that must be satisfied
by the exact solutions of the characteristic value problem
(see eqs.[24] in LM).

In virtue of equations (89), we can also combine equations
(91) and (92) in the manner
$$
{ \langle { A { \boldeta } ( m ), { \boldeta } ( n ) } \rangle}_2
= \sum_{j=1}^Q (-1)^j \alpha (j,m) \alpha (j,n) = \delta_{mn}. 
\eqno\stepeq
$$
In other words, the constants $\alpha (j,n)$ are normalized in the manner
$$
\sum_{j=1}^Q (-1)^j [ \alpha (j,n) ]^2  = 1 . 
\eqno\stepeq
$$

\subsection{Identification of modes in different approximations}

We shall distinguish solutions of equations (87) obtained in different
approximations with the aid of subscripts.  Thus, let
$\omega_a (n)$, $a_a (j,n)$, $\alpha_a (j,n)$, and 
$\boldeta_a (n)$ denote the quantities involved in the representation
of the $n${th} characteristic solution of equations (87) in the $a${th}
approximation, where $a = k_{\rm max} + 1$.  Accordingly, we write
equation (88) for the $n${th} characteristic solution
for the Lagrangian displacement in the $a${th} approximation as
$$
{ \boldeta }_a (n) = \sum_{j=1}^{N_a} a_a (j,n) { \bbeta }_a ( j ) ,
\eqno\stepeq
$$
where $N_a = (k_{\rm max} + 1)(k_{\rm max} + 3) = a(a+2)$.
The basis vectors $\bbeta_a (j)$ are labelled here with a subscript $a$,
because the sets of basis vectors constructed in Section 4.3 are
different in the different approximations.  The number of characteristic
solutions of equation (87) in the $a${th} approximation is equal to
$Q_a = (k_{\rm max} + 1)(k_{\rm max} + 2) = a(a+1)$

A useful comparison of the characteristic solutions $\boldeta_a (n)$ and
$\boldeta_b (m)$ $(a>b)$ is provided by the inner product
$$
I_{ab} (n,m) =
{ \langle { A { \boldeta }_a ( n ), { \boldeta }_b ( m ) } \rangle}_2
= \sum_{j=1}^{N_a} \sum_{k=1}^{N_b} a_a (j,n) a_b (k,m)
{ \langle { A { \bbeta }_a ( j ), { \bbeta }_b ( k ) } \rangle}_2 .
\eqno\stepeq
$$
The values of $|I_{ab} (n,m)|$ provide criteria for identifying
characteristic solutions of equations (87) that represent the same mode
in different approximations.  Note that we take the modulus of the inner
product here, because the phases of the solutions $\boldeta_a (n)$ and
$\boldeta_b (m)$ need not be the same.  According to equation (24) in LM,
the Lagrangian displacements of different modes are mutually orthogonal.
If $\boldeta_a (n)$ and $\boldeta_b (m)$ were accurate representations of
the Lagrangian displacement of the same mode, then we would have
$|I_{ab} (n,m)| = 1$.  Otherwise, if $\boldeta_a (n)$ and
$\boldeta_b (m)$ were accurate representations of the Lagrangian
displacements of different modes, then we would have
$|I_{ab} (n,m)| = 0$.  The characteristic solutions of equations (87)
give only approximate representations of the Lagrangian displacements
of modes, in general, so the values of $|I_{ab} (n,m)|$ should lie
between 0 and 1 in practice.

Consider now the evaluation of the inner products
${ \langle { A { \bbeta }_a ( j ), { \bbeta }_b ( k ) } \rangle}_2$
on the right-hand side of equation (96).
The prescription for the construction of basis vectors described
in Section 4.3 (see particularly the ordering of the vector polynomials
in equations [52] and [53] there) implies that
$$
{ \bbeta }_b ( k ) = { \bbeta }_a ( k )  \quad (k=1,2,\dots,Q_b ) .
\eqno\stepeq
$$
Therefore, we can reduce equation (96) to the form
$$
I_{ab} (n,m) =
\sum_{k=1}^{Q_b} a_a (k,n) a_b (k,m) + 
\sum_{j= Q_b + 1}^{N_a} \sum_{k= Q_b + 1}^{N_b} a_a (j,n) a_b (k,m)
{ \langle { A { \bbeta }_a ( j ), { \bbeta }_b ( k ) } \rangle}_2 
\eqno\stepeq
$$
with the aid of the orthogonality relations governing the basis
vectors (see eqs. [64]).  Substituting from equations (57) for
${ \bbeta }_b ( k )$, and again making use of equations (64)
and (97), we verify that
$$
{ \langle { A { \bbeta }_a ( j ), { \bbeta }_b ( k ) } \rangle}_2
= N_b (k)
{ \langle { A { \bbeta }_a ( j ), { \bpi }_b ( k ) } \rangle}_2
\quad (Q_b < j \le N_a;Q_b < k \le N_b ) .
\eqno\stepeq
$$
It also follows from equations (52) and (53) that
$$
\bpi_b ( k ) = \bpi_a ( l ) \quad (k > Q_b ; l = k + Q_a - Q_b > Q_a ) .
\eqno\stepeq
$$
We also have
$$
N_b ( k ) = N_a ( l ) = 1 +{\rm i} \quad (k > Q_b ; l = k + Q_a - Q_b > Q_a )
\eqno\stepeq
$$
in virtue of equations (76).  We can therefore reduce the right-hand
side of equation (99) in the manner
$$
\eqalign{
{ \langle { A { \bbeta }_a ( j ), { \bbeta }_b ( k ) } \rangle}_2
&= N_a (l)
{ \langle { A { \bbeta }_a ( j ), { \bpi }_a ( l ) } \rangle}_2 \cr
&= - N_a (l) \sum_{r=1}^{Q_a} c_a(l,r) 
{ \langle { A { \bbeta }_a ( j ), { \bbeta }_a ( r ) } \rangle}_2
\quad (Q_b < j \le N_a; Q_a < l = k + Q_a - Q_b \le N_b + Q_a - Q_b) ,
 \cr } \eqno\stepeq
$$
where we have eliminated ${ \bpi }_a ( l )$ with the aid of equations (57)
and we have suppressed terms that vanish identically with the aid
equations (24), (64), and (75). Upon substituting from equation (102)
for ${ \langle { A { \bbeta }_a ( j ), { \bbeta }_b ( k ) } \rangle}_2$
on the right-hand side of equation (98) and making further use of equations
(24) and (64) in the evaluation of the inner products
${ \langle { A { \bbeta }_a ( j ), { \bbeta }_a ( r ) } \rangle}_2$,
we finally obtain
$$
I_{ab} (n,m) =
\sum_{k=1}^{Q_b} a_a (k,n) a_b (k,m) - 
\sum_{j= Q_b + 1}^{Q_a} \sum_{l= Q_a + 1}^{N_b + Q_a - Q_b}
a_a (j,n) a_b (l - Q_a + Q_b,m) N_a (l) c_a (l,j) .
\eqno\stepeq
$$

\section{Numerical examples}

In this section, we illustrate the matrix method with a calculation
of modes of radial oscillation in a family of spherically symmetric
models of galaxies.  In each member of the family, the unperturbed
distribution function is a function of the energy alone, and the
unperturbed density is of the form
$$
\rho_0 (r) = { { 3M_0 {r_c}^2 } \over { 4 \upi ( {r_c}^2 + r^2 )^{5/2} } }
- { { 3M_0 {r_c}^2 } \over { 4 \upi ( {r_c}^2 + R^2 )^{5/2} } } \qquad
(r \le R) ,  \eqno\stepeq
$$
where $R$ is the radius of the configuration, the constant $ r_c $
is of the nature of a core radius, and the constant $M_0$ is a
certain characteristic mass.  It follows from equation (104)
that the characteristic mass is determined in terms
of the central density $\rho_0 (0)$, the radius $R$, and the core
radius $r_c$ in the manner
$$
M_0 = { {4 \upi} \over 3} \rho_0 (0) R^3 {x_c}^3
{\left[ 1 - { { {x_c}^5 } \over { ({x_c}^2 + 1 )^{5/2} } } \right]}^{-1} ,
\eqno\stepeq
$$
where $x_c = {r_c}/R$.  We can evaluate the mass of the system
interior to a spherical surface of radius $r$ as
$$
M_0 (r) = 4 \upi \int_0^r \rho_0 (r) r^2 {\rm d}r
= { { M_0 r^3 } \over { ( {r_c}^2 + r^2 )^{3/2} } }
- { { M_0 {r_c}^2 r^3 } \over { ( {r_c}^2 + R^2 )^{5/2} } } ,
\eqno\stepeq
$$
and we can accordingly evaluate the gravitational acceleration in the
unperturbed system with the aid of Newton\rq{s} theorem.  In the
limit $ R \rightarrow \infty $, the members of this family of models
tend to Plummer{\rq}s model.

We make use of $R$, $ \rho_0 (0) $, and
$ [ \upi G \rho_0 (0) ]^{-1/2} $ as the units of length, density,
and time, respectively.  In this system of units, the models
form a one-parameter sequence.  The dimensionless core radius
$x_c = {r_c}/R$ is a convenient
parameter with which to distinguish the members of the sequence.  
Alternatively, we can make use of the central concentration,
defined here as the ratio $C = \rho_0 (0)/{\bar \rho_0}$ of the
central density to the mean density of a configuration, as the
parameter which distinguishes models along the sequence.
With the aid of equations (105) and (106), it is a straight-forward
calculation to verify that
$$
C \equiv { { \rho_0 (0) } \over { {\bar \rho_0} } }
= { {4 \upi} \over 3 } \rho_0 (0) R^3
{ \left[ {4 \upi \int_0^R \rho_0 (r) r^2 {\rm d}r } \right] }^{-1}
= {1 \over {x_c}^3 } [ ({x_c}^2 + 1 )^{5/2} - {x_c}^5 ] .
\eqno\stepeq
$$

The solutions of equations (87) occur in pairs, the frequencies
of each pair being equal in magnitude but opposite in algebraic signs,
so it will suffice to present results of the present application of
the matrix method only for modes with positive frequencies.
Moreover, we consider only modes which we can identify in both
the third and fourth approximations.  As is explained at the end of
Section 5, we make use of the values of the inner products
$I_{43} (n,m)$ as criteria for the identification of such modes.
For a given unperturbed system, we have
12 solutions of equations (87) in the third approximation
$(k_{\rm max} = 2, Q = 12)$, and 20 solutions in the fourth approximation
$(k_{\rm max} = 3, Q = 20)$.  Thus, we
can expect to identify up to six modes with positive frequencies
by comparing solutions of equations (87) in the third and fourth
approximations.  For this purpose  we have
evaluated $|I_{43} (n,m)|$ for all pairs of values $(n,m)$
$(n=1,2,\dots,20;m=1,2,\dots,12)$, but we have arbitrarily selected
for further investigation only those combinations in which
$|I_{43} (n,m)| \ge 0.4$.
Making similar use of the inner products $|I_{42} (n,m)|$ and $|I_{32} (n,m)|$
as criteria, we can also identify solutions of equations (87) in
the second approximation which represent up to three of the
modes identified in the third and fourth approximations.

\begintable*{1}
  \caption{{\bf Table 1.} Identification of modes of radial oscillation
     in a system in which $C = 11.335$ $(x_c = 0.55)$.}
  \halign{#\hfil & \quad \hfil#\hfil\quad & \hfil#\hfil\quad &
      \hfil#\hfil\quad & \hfil#\hfil\quad & \hfil#\hfil\quad &
       \hfil#\hfil\quad & \hfil#\hfil\quad & \hfil#\hfil\quad &
        \hfil#\hfil\cr
  Mode  & $n$ & $m$ & $l$ & $|I_{43} (n,m)|$ & $|I_{42} (n,l)|$ & $|I_{32} (m,l)|$ &
        $\omega_4 (n)$ & $\omega_3 (m)$ & $\omega_2 (l)$ \cr
  R1 & 19 & 9 & 5 & 0.9999 & 0.9994 & 0.9995 & 0.6143 & 0.6144 & 0.6162 \cr
  R2 & 20 & 12 & {} & 0.9088 & {} & {} & 0.8628 & 1.1280 & {} \cr
  R3 & 18 & 11 & 6 & 0.9243 & 0.8448 & 0.7429 & 1.3000 & 1.3510 & 1.4180 \cr
  R4 & 17 & 10 & {} & 0.9938 & {} & {} & 1.6360 & 1.6460 & {} \cr
  R6 & 15 & 8 & 4 & 0.8928 & 0.5608 & 0.6499 & 2.4700 & 2.6310 & 2.3510 \cr
  R9 & 4 & 2 & {} & 0.8666 & {} & {} & 4.1300 & 4.2430 & {} \cr }
\endtable

Table 1 illustrates the identification of six modes in an unperturbed
configuration in which  $x_c \equiv r_c / R = 0.55$ and
$C \equiv \rho_0 (0)/{\bar \rho_0} = 11.335$ (see eq. [107]).
In the first column there, we label modes in terms of their
characteristic frequencies in the fourth approximation.
We arrange the positive solutions of equations (87) for the
frequencies in the fourth approximation in a sequence in the order of
increasing values, and we identify the $n${th} frequency in that sequence
and the mode to which it belongs with the label \lq{R$n$}\rq.
The integers $l$, $m$, and $n$ in the next three columns of Table 1
identify solutions of equations (87)
in the second, third, and fourth approximations, respectively, with
labels which were assigned by the LANPAK subroutines used in the
solution of equations (87).  The inner products $I_{43} (n,m)$,
$I_{42} (n,l)$, and $I_{32} (m,l)$ have been evaluated with the aid
of equation (103).  In the last three columns of the table,
$\omega_2 (l)$, $\omega_3 (m)$, and $\omega_4 (n)$ denote solutions
for the characteristic frequencies in the second, third, and fourth
approximations, respectively.

The solutions listed in Table 1
and their counterparts with negative frequencies are all of the
cases in which $|I_{43} (n,m)| \ge 0.4$.  The result that the listed
values of $|I_{43} (n,m)|$ (values which also apply to the corresponding cases
with negative frequencies) exceed this limit by such a wide margin and
approach unity so closely indicates that this comparison of the solutions
of equations (87) in the third and fourth approximations is a
reliable basis for the identification of modes.
The trends in the values of $|I_{32} (m,l)|$, $|I_{42} (n,l)|$,
and $|I_{43} (n,m)|$ for the modes which are also represented in the
second approximation indicate that the sequence of approximations
represented in Table 1 tends to converge.  The values of the characteristic
frequencies listed in Table 1 in the different approximations also indicate that
the identification of modes in the different approximations is reliable and
that the sequence of approximations tends to converge.

For each of the modes represented in Table 1, we plot in Figures 1-6 the
solutions in the second, third, and fourth approximations
for the amplitudes $\alpha (k)$ and for the quantity
$$
M_1 (r,t) = M_1 (r,0) \exp (-{\rm i} \omega t)
= \int_{|\bmath{x}|<r} \rho_1 (\bmath{x},t) {\rm d} \bmath{x}
= - \int_{|\bmath{x}|<r}
\left[ {\upartial \over {\upartial \bmath{x} } } \cdot
(\rho_0 \bxi ) \right] {\rm d} \bmath{x}
= - 4 \upi r^2 \rho_0 (r) {  \bmath{x}  \over r } \cdot
\bxi (\bmath{x},t) , \eqno\stepeq
$$
which is the Eulerian perturbation of the mass interior to a spherical
surface of radius $r$.  In writing this expression for $M_1 (r,t)$, we
have represented the Eulerian perturbation of the density
$\rho_1 (\bmath{x},t)$ in terms of the hydrodynamical Lagrangian
displacement $\bxi (\bmath{x},t)$ with the aid of the relations
$$
\rho_1 (\bmath{x},t) = - {\upartial \over {\upartial \bmath{x} } }
\cdot (\rho_0 \bxi )
\quad \hbox{and} \quad
\rho_0 (r) \bxi (\bmath{x},t) =
m_* \int \Delta \bmath{x} (\bmath{x},\bmath{v},t) f_0 (\bmath{x},\bmath{v})
{\rm d}\bmath{v}  \eqno\stepeq
$$
(Vandervoort 1983).  In the second of equations (109), the integration
extends over the region of the velocity space accessible to stars in the
neighborhood of the point $\bmath{x}$.  In order to evaluate $\bxi (\bmath{x},t)$,
we express the Lagrangian displacement $\boldeta$ as a superposition of
the vector monomials
${\bmu}_{\sigma}^{\tau} ({\bmath{z}}_{+1},{\bmath{z}}_{-1} ;m,n,p)$
with the aid of equations (55), (57), and (88), we extract from that
superposition an expression for $\Delta \bmath{x} (\bmath{x},\bmath{v},t)$,
and we express the result of integrating $\Delta \bmath{x} (\bmath{x},\bmath{v},t)$
over the velocity space as a superposition of the moments $G(r, \mu)$ of
the velocity distribution described in Appendix E.
\beginfigure*{1}
\vskip 67mm
\caption{{\bf Figure 1.} Representations of the mode R1 in three
approximations in a system in which $C = 11.335$ $(x_c = 0.55)$.
This is the fundamental mode of radial oscillation.
(a) Amplitudes $\alpha (k)$ in the representation of the perturbation.
(b) Dependence of the Eulerian perturbation of the mass variable
Re$[M_1 (r,0)]$ on the distance $r$ from the centre of the unperturbed
system. }
\endfigure

It happens that we can
choose the phase of the solution for each mode so that the solutions
for $\alpha (k)$ are real; we adopt that choice in all of the
calculations described below.  A comparison of the amplitudes $\alpha (k)$
in different approximations is a direct comparison of solutions of the
characteristic value problem in the reduced matrix representation
described by equations (87).  Implicitly, this is a comparison of
Lagrangian displacements in the different approximations.  With the phase
of the mode chosen so that the coefficients $\alpha (k)$ are
real, it turns out that $M_1 (r,0)$ is a complex quantity
with equal real and imaginary parts.  It will therefore suffice
to plot the real part of $M_1 (r,0)$ in what follows.  A comparison of
the solutions for $Re [M_1 (r,0)]$ in the different approximations is
a comparison of the different approximations for the
structures of perturbations in the configuration space.

The mode R1 listed in Table 1 is the fundamental mode of radial
oscillation.  Evidently, the fundamental mode is represented very
accurately in these calculations.  The plots of the solutions for
$\alpha (k)$ and Re$[M_1 (r,0)]$ are displayed in Figure 1.  The
mode is already well represented in the second approximation, and
the additional amplitudes $\alpha (k)$ that appear in the third and
fourth approximations are almost negligible.  The solutions for
Re$[M_1 (r,0)]$ converge rapidly.

The mode R2 is not represented very accurately in these calculations.
According to Table 1, the discrepancy in the values of the frequency
in the third and fourth approximations is substantial.  In Figure 2a,
the amplitudes $\alpha (k)$ $(12 < k \le 20)$ in the fourth approximation
are not negligible, so the representations of the mode in the
third and fourth approximations must be significantly different.
The difference is manifest in the plots of Re$[M_1 (r,0)]$ in
Figure 2b.
\beginfigure*{2}
\vskip 67mm
\caption{{\bf Figure 2.} Representations of the mode R2 in two
approximations in a system in which $C = 11.335$ $(x_c = 0.55)$.
(a) Amplitudes $\alpha (k)$ in the representation of the perturbation.
(b) Dependence of the Eulerian perturbation of the mass variable
Re$[M_1 (r,0)]$ on the distance $r$ from the centre of the unperturbed
system. }
\endfigure
\beginfigure*{3}
\vskip 67mm
\caption{{\bf Figure 3.} Representations of the mode R3 in three
approximations in a system in which $C = 11.335$ $(x_c = 0.55)$.
(a) Amplitudes $\alpha (k)$ in the representation of the perturbation.
(b) Dependence of the Eulerian perturbation of the mass variable
Re$[M_1 (r,0)]$ on the distance $r$ from the centre of the unperturbed
system. }
\endfigure
\beginfigure*{4}
\vskip 67mm
\caption{{\bf Figure 4.} Representations of the mode R4 in two
approximations in a system in which $C = 11.335$ $(x_c = 0.55)$.
(a) Amplitudes $\alpha (k)$ in the representation of the perturbation.
(b) Dependence of the Eulerian perturbation of the mass variable
Re$[M_1 (r,0)]$ on the distance $r$ from the centre of the unperturbed
system. }
\endfigure
\beginfigure*{5}
\vskip 67mm
\caption{{\bf Figure 5.} Representations of the mode R6 in three
approximations in a system in which $C = 11.335$ $(x_c = 0.55)$.
(a) Amplitudes $\alpha (k)$ in the representation of the perturbation.
(b) Dependence of the Eulerian perturbation of the mass variable
Re$[M_1 (r,0)]$ on the distance $r$ from the centre of the unperturbed
system. }
\endfigure
\beginfigure*{6}
\vskip 67mm
\caption{{\bf Figure 6.} Representations of the mode R9 in two
approximations in a system in which $C = 11.335$ $(x_c = 0.55)$.
(a) Amplitudes $\alpha (k)$ in the representation of the perturbation.
(b) Dependence of the Eulerian perturbation of the mass variable
Re$[M_1 (r,0)]$ on the distance $r$ from the centre of the unperturbed
system. }
\endfigure

The convergence of the representations of mode R3 in the three
approximations that is already evident in Table 1 is also evident
in Figure 3.  The amplitudes $\alpha (k)$ in the three approximations
are in relatively good agreement, and, in the fourth approximation
the amplitudes $\alpha (k)$ $(12 < k \le 20)$ are relatively small
if not quite negligible.  The plots of Re$[M_1 (r,0)]$ show a
satisfactory convergence.

The representations of the mode R4 in Table 1 and Figure 4 appear to
be quite accurate.  The solutions for the amplitudes $\alpha (k)$
$(1 \le k \le 12)$ in the third and fourth approximations agree very well,
and the values of $\alpha (k)$ $(12 < k \le 20)$ in the fourth approximation
are quite small.  The runs of Re$[M_1 (r,0)]$ in the two approximations
are remarkably similar.

The values of $|I_{32} (m,l)|$ and $|I_{42} (n,l)|$ listed in Table 1
for the mode R6 indicated that that mode is not accurately represented
in the second approximation.  In Figure 5(a), the amplitudes
$\alpha (k)$ in the second approximation do not agree very well
with their counterparts in the third and fourth approximations,
and the amplitudes $\alpha (k)$ $(k > 6)$ are substantial in the
third and fourth approximations.  Likewise, in the runs of
Re$[M_1 (r,0)]$ in Figure 5(b), the second approximation does not
provide an accurate representation of the mode.  Nevertheless,
the agreement of the representations of the mode R6 in the third
and fourth approximations is quite good.

In the plots of $\alpha (k)$ and Re$[M_1 (r,0)]$ in Figure 6, it appears
that the identification of the mode R9 in Table 1 is reliable and
that the representation of the mode is relatively accurate.

The preceding discussion illustrates the matrix method and identifies
six of the lowest modes of radial oscillation in a system with
a relatively modest central concentration $(C \equiv \rho_0 (0)/{\bar \rho_0}
= 11.335; x_c \equiv r_c / R = 0.55)$.  Figure 7 shows the
dependence of the characteristic periods $P = 2 \pi / \omega$ on the central
concentration $C$ for four modes which can be identified in both the third
and fourth approximations and traced
over a range of substantially greater central concentrations.  Values
of the periods derived in the second, third, and fourth approximations
are plotted as open triangles, circles, and squares, respectively.
We have chosen to plot the periods instead of the frequencies
in Figure 7 in order to facilitate
comparisons of the different approximations in systems with higher
central concentrations.  The criterion adopted here for the identification
of a mode in the third and fourth approximations is the condition
$|I_{43} (n,m)| \ge 0.75$.  When this criterion is satisfied, points
representing the periods in the third and fourth approximations are
plotted as large circles and large squares respectively.  The smaller
circles and squares represent extrapolations or interpolations to
values of the central concentrations at which the values of
$|I_{43} (n,m)|$ are less than 0.75.  Two of the modes
presented in Figure 7 are also represented in the second approximation,
and those results are plotted as large triangles or small triangles
according as $|I_{42} (n,l)| \ge 0.75$ or $|I_{42} (n,l)| < 0.75$.

\beginfigure*{7}
\vskip 136mm
\caption{{\bf Figure 7.} Dependence of the characteristic period
$P = 2 \pi / \omega$ on the central
concentration C for four modes of radial oscillation.  Open triangles,
circles, and squares represent the second, third, and fourth
approximations, respectively.  The different sizes of the symbols
are explained in the text.  Filled circles represent the results
of N-body solutions of the Lagrangian perturbation equations. }
\endfigure
In the present calculations, it has not been possible to obtain
accurate solutions of equations (87) in the fourth approximation
for unperturbed systems in which $C > 1025$ $(x_c < 0.1)$.  The
failure of these calculations for the most centrally concentrated
systems considered was traced to the accumulation of round-off
errors in the numerical solution of equations (61)-(63) for the
quantities $N(q)$ and $c(q,r)$.

The results displayed in Figure 7(a) indicate that the present
representations of the fundamental mode, the mode R1, remain quite
accurate in systems with central concentrations up to values at least
of the order of 10,000.  The behavior of the characteristic frequency
in the fourth approximation is described in the most centrally
concentrated systems by the formula
$$
\omega \approx 3.5 ( \upi G {\bar \rho_0} )^{1/2} \quad (C >> 100) ,
\eqno\stepeq
$$
where ${\bar \rho_0}$ is the mean density of the system.  For the
fundamental mode in systems in which $C = 11.34$ $(x_c = 0.55)$
and in which $C = 1025$ $(x_c = 0.1)$ plots of $\alpha (k)$ and
Re$[M_1 (r,0)]$ are shown in Figures 1 and 8, respectively.  Whereas
the second, third, and fourth approximations are substantially the
same in the less centrally concentrated system represented in Figure 1,
the three approximations are rather different in the more centrally
concentrated system represented in Figure 8.  Nevertheless, the
sequence of approximations does appear to be converging to an
accurate representation of the mode in the latter case.  The node near
the origin in the run of Re$[M_1 (r,0)]$ in Figure 8(b) is unexpected.
That feature need not be real; it could be a result of the
accumulation of numerical errors mentioned above or it could be
an artifact of the present method of approximation.  That
this feature is absent in the second approximation, present
in the third approximation, and almost absent in the fourth
approximation suggest that it might be an artifact.  The feature
does not appear at lower central concentrations.

Figure 7(a)
includes a comparison of the present results with values of the
period of the fundamental mode calculated for the same unperturbed
systems with the aid of an N-body method for the solution of the
Lagrangian perturbation equations (Vandervoort 1999).  The N-body
results, which are represented by filled circles in Figure 7(a), were
obtained before the start of the present work on the matrix method.
A mode is excited in such N-body experiments by imposing suitable
initial conditions on the Lagrangian displacements 
$(\Delta \bmath{x},\Delta \bmath{v})$ of the particles.  In the
experiments represented in Figure 7(a), suitable initial conditions were
found by a process of trial and error in which corrections were
applied to virial estimates of the Lagrangian displacements
(Vandervoort 1983,1999).  Those efforts to excite the fundamental
mode without exciting other perturbations significantly were
successful only for systems in which the central concentrations are
sufficiently low $(C \le 137.8 ; x_c \ge 0.2)$.  In future work, it will
be of interest to make use of Lagrangian displacements derived with
the aid of the matrix method as initial conditions with which to
excite the fundamental mode and other modes in N-body experiments
on systems covering a wider range of central concentrations.

In Figure 7(b), it appears that the identification of the mode R2
is quite firm over the full range of central concentrations spanned
by the plotted points.  Even the smaller points plotted at
$C = $ 17.20, 22.49, and 31.01 represent cases in which
$|I_{43} (n,l)| > 0.7$.  Figure 9 shows plots of $\alpha (k)$ and
Re$[M_1 (r,0)]$ for the mode R2 in the case that $C = 1025$ $(x_c = 0.1)$.
The plots of $\alpha (k)$ in the third and fourth approximations
appear to be quite different; nevertheless the runs of Re$[M_1 (r,0)]$
in the two approximations agree rather well.  The node that appears
near the origin in the run of Re$[M_1 (r,0)]$ in the fourth
approximation seems to be an artifact of the approximation like
the similar feature mentioned above in the case of the fundamental
mode.  The fourth approximation to the mode R2 does not have such a
feature when $C = 137.8$ $(x_c = 0.2)$.  A comparison of Figure 9(b)
with Figure 2(b) indicates that the mode R2 has significantly
different structures in systems of high and low central concentrations.
The run of Re$[M_1 (r,0)]$ has two interior nodes at $C = 11.34$
$(x_c = 0.55)$ but only one interior node at $C = 1025$ $(x_c = 0.1)$.
The transition from two nodes to one node occurs at $C \approx 13.59$
$(x_c \approx 0.5)$ in the third approximation and at $C \approx 17.20$
$(x_c \approx 0.45)$ in the fourth approximation.
\beginfigure*{8}
\vskip 67mm
\caption{{\bf Figure 8.} Representations of the mode R1 in three
approximations in a system in which $C = 1025$ $(x_c = 0.1)$.
This is the fundamental mode of radial oscillation.
(a) Amplitudes $\alpha (k)$ in the representation of the perturbation.
(b) Dependence of the Eulerian perturbation of the mass variable
Re$[M_1 (r,0)]$ on the distance $r$ from the centre of the unperturbed
system. }
\endfigure
\beginfigure*{9}
\vskip 67mm
\caption{{\bf Figure 9.} Representations of the mode R2 in two
approximations in a system in which $C = 1025$ $(x_c = 0.1)$.
(a) Amplitudes $\alpha (k)$ in the representation of the perturbation.
(b) Dependence of the Eulerian perturbation of the mass variable
Re$[M_1 (r,0)]$ on the distance $r$ from the centre of the unperturbed
system. }
\endfigure

Although it was shown above that, for the system in which $C = 11.34$
$(x_c = 0.55)$, the identification of the mode R4 is reliable and the
representation of the mode is accurate, the identification and
representation of the mode becomes problematic at higher central
concentrations.  In the representation of the mode in Figure 7(c), the
runs of two of the solutions for the characteristic frequency in the
third approximation are shown.  The application of the criterion
$|I_{43} (n,m)| \ge 0.75$ implies that the lower of the two curves
represents the mode R4 in the cases that $C \le 22.49$
$(x_c \ge 0.4)$ whereas the higher curve represents the mode
in the cases that $31.01 \le C \le 137.8$ $(0.35 \ge x_c \ge 0.2)$.
At central concentrations exceeding 137.8, we find no solution of
equation (87) in the third approximation that can be identified
convincingly as a representation of the mode R4.  (Of course the
nomenclature adopted in order to label modes implies that the
mode R4 is correctly represented in the fourth approximation by
definition.)   Plots of $\alpha (k)$
and Re$[M_1 (r,0)]$ are shown in Figure 10 for the mode R4 in the
case that $C = 137.8$ $(x_c = 0.2)$.  These results in the third
and fourth approximations agree well enough to support the identification
of the mode in both approximations but not well enough to show that
we have an accurate solution for the mode in this case.  And we
have no indication of the accuracy with which the mode is represented
in the fourth approximation at greater central concentrations.
\beginfigure*{10}
\vskip 67mm
\caption{{\bf Figure 10.} Representations of the mode R2 in two
approximations in a system in which $C = 137.8$ $(x_c = 0.2)$.
(a) Amplitudes $\alpha (k)$ in the representation of the perturbation.
(b) Dependence of the Eulerian perturbation of the mass variable
Re$[M_1 (r,0)]$ on the distance $r$ from the centre of the unperturbed
system. }
\endfigure
\beginfigure*{11}
\vskip 67mm
\caption{{\bf Figure 11.} Representations of the mode R8 in two
approximations in a system in which $C = 1025$ $(x_c = 0.1)$.
(a) Amplitudes $\alpha (k)$ in the representation of the perturbation.
(b) Dependence of the Eulerian perturbation of the mass variable
Re$[M_1 (r,0)]$ on the distance $r$ from the centre of the unperturbed
system. }
\endfigure

As is shown in Figure 7(d), the mode R8 can be identified in accordance
with the criterion $|I_{43} (n,m)| \ge 0.75$ only in centrally
concentrated systems.  The smaller circles and squares represent points
in the third and fourth approximations, respectively, at which
$0.55 < |I_{43} (n,m)| < 0.75$.  Likewise, the small triangles represent
points on the run of the period in the second approximation at which
$0.55 < |I_{42} (n,l)| < 0.75$.   Plots of $\alpha (k)$
and Re$[M_1 (r,0)]$ are shown in Figure 11 for the mode R8 in the
case that $C = 1025$ $(x_c = 0.1)$.  The solutions for the
amplitudes $\alpha (k)$ in the three approximations agree quite well
in Figure 11(a), but the amplitudes that are added at higher values
of $k$ in the third and fourth approximations are significant.
The three approximations for the run of Re$[M_1 (r,0)]$ agree
quite well in Figure 11(b).  It should be noted, however, that
the three approximations do not agree on the number of nodes in
the outer regions of the system.  The hydrodynamical
Lagrangian displacement is badly determined in the outer regions
of a centrally concentrated system because there is very little
mass there.  In particular, the (two or) three approximations often
disagree on the number of interior nodes in the run of the hydrodynamical
Lagrangian displacement near the outer boundary of the unperturbed system.  

\section{The continuous spectrum of normal modes}

In all of the cases described in Section 6, the modes that have been
identified have frequencies which lie within the continuous spectrum of
the frequencies of the radial components of the stellar orbits.
Specifically, the frequencies of the modes considered here lie within the
interval
$$
{ {GM_0 (R)} \over R^3 } \le \omega^2 \le { { 16 \upi G\rho_0} \over 3} .
\eqno\stepeq
$$
The lower limit in this inequality is the Keplerian frequency of an orbit
at the outer edge of the system, whereas the upper limit is the
frequency of the radial component of a (small) simple-harmonic
oscillation of a star about the centre of the system.  Inequality (111)
implies that the modes identified here lie within the continuous spectrum
of radial modes that is to be expected in the systems considered.  

In general, the normal modes of oscillation of a galaxy include a set
of modes with a discrete spectrum of real and complex frequencies and
a set of modes with a continuous spectrum of real frequencies.  Modes
of instability would belong to the discrete spectrum.  The frequencies
belonging to the continuous spectrum satisfy resonance conditions with
the frequencies of the unperturbed stellar orbits in the system.  In
particular, frequencies satisfying inequality (111) would satisfy
appropriate resonance conditions for a continuous spectrum of
radial modes in the systems considered in the present work.
Examples of systems in which the normal modes include both discrete and
continuous spectra of frequencies are described by Fridman \&
Polyachenko (1984; see particularly Chapter II, Section 7 and Chapter
III, Sections 3.4.1 \& 6.1.3).  Mathur (1990; see also Weinberg 1991b)
has investigated conditions under which the normal modes of a system
include a discrete spectrum of real frequencies which lie outside the
continuous spectrum.

The question arises in the present work as to how discrete modes can
appear within the continuous spectrum of modes.  The probable
explanation is that such discrete modes are isolated members of
the continuous spectrum with frequencies which \lq{accidently}\rq\/
satisfy the characteristic equation.

This interpretation is supported by an investigation of the
solutions of the Lagrangian perturbation equations for the complete
spectra of normal modes in stellar systems (Vandervoort 2001).  That
investigation is based on a matrix method for the study of perturbations
which is rather different from the method described in this paper but
similar to the matrix method of Kalnajs (1977).  For our present
purposes, the principal results are the following.  For the discrete
spectrum of modes in a given system, there is a characteristic
equation for the determination of the frequencies. In general,
however, there is no characteristic equation governing frequencies
in the continuous spectrum.  On the other hand, there are
exceptional frequencies in the continuous spectrum which do
satisfy the characteristic equation for discrete modes.  Although the
details of that work must be left to a later publication, the circumstances
in which isolated frequencies in the continuous spectrum can satisfy
the characteristic equation can be illustrated with the aid of well
known results in the theory of plasma oscillations.

In the second matrix formulation of Vandervoort (2001), the modes of oscillation
of a galaxy are recognized as counterparts of the van Kampen modes
of oscillation in a homogeneous plasma (van Kampen 1955,
1957; Case 1959).  For the sake of simplicity, we consider a model of
a plasma which consists of an electron gas of uniform density in the presence
of a fixed, uniform background of positive charge. A van Kampen mode
is an electrostatic plane wave of wave number $k$ and frequency $\omega$,
say.  The simplest example is obtained for a one-dimensional representation
in which the electrostatic field is parallel to the direction of the wave,
and functions of the velocity of an electron are integrated over the
components of the velocity perpendicular to the direction of the wave.
The Eulerian perturbation of the distribution function of the electrons and the
electrostatic field are determined by solving Vlasov{\rq}s equation
(i.e., the linearized, collisionless Boltzmann equation) and Poisson{\rq}s
equation simultaneously.  The Eulerian perturbation of the
distribution function is of the form
$f_1 (x,v,t) = g(v) {\rm exp}[{\rm i}(kx- \omega t)]$ where
the $x-$direction is the direction of the wave and the amplitude $g(v)$
is a function of the $x-$component $v$ of the velocity.  The solution for
$g(v)$ is of the form
$$
g(v) = {u_P}^2 {\rm P} {{{\rm d} f_0 (v)} \over {{\rm d} v }} { 1 \over { u - v } }
+ \lambda \delta (u - v) ,
\eqno\stepeq
$$
where $f_0 (v)$ is the distribution function of the unperturbed electron gas,
$u = \omega / k$ is the phase velocity of the wave, $u_P = \omega_P / k$ is
a characteristic velocity derived from the plasma frequency $\omega_P$, the symbol
${\rm P}$ denotes the Cauchy principal value, $\delta (u - v)$ is Dirac{\rq}s
delta function, and $\lambda$ is a constant to be determined.  Applying the
normalization condition $ \int g(v) {\rm d} v = 1 $ to equation (112), we obtain
$$
\lambda = 1 - {u_P}^2 {\rm P} \int_{- \infty}^{\infty}
{{{\rm d} f_0 (v)} \over {{\rm d} v }} { 1 \over { u - v } } {\rm d} v .
\eqno\stepeq
$$
For an arbitrary choice of the value of $u = \omega / k$, in general,
equation (113) determines the value of $\lambda$.  Thus, in general, there
is no dispersion relation (characteristic equation) for the determination
of the frequencies $\omega$ of the van Kampen modes in the continuous spectrum.
For isolated frequencies, however, it can happen that $\lambda = 0$.
Such frequencies can be determined by setting the right-hand side of equation
(113) equal to zero.  The condition that the right-hand side of equation
(113) vanish is the dispersion relation originally derived by Vlasov (1945)
for plasma oscillations.  It has been remarked (Bohm \& Gross 1949,
Bernstein, Greene \& Kruskal 1957, Jackson 1960) that Vlasov's
dispersion relation picks out the frequencies of waves in which there
are no electrons trapped in the electrostatic field.

In precisely the same way, isolated members of the continuous spectrum
of modes in a galaxy are found to satisfy the characteristic equation
for discrete modes.  It is commonly expected that such perturbations
would suffer Landau damping.  However, Landau damping should not occur
in individual modes in a spherical system in which the unperturbed
distribution function depends only on the energy of a star.  According
to a theorem of Antonov (1961; Fridman \& Polyachenko 1984; Binney \&
Tremaine 1987; Palmer 1994), such a system will be stable if the
distribution function
is a monotonically decreasing function of the energy, as it would be
in the systems considered in Section 6.  Moreover, a lemma proved in LM
shows that the frequencies of normal modes in a stellar system occur, in
general, in complex-conjugate pairs.  Thus, the frequencies of the modes
considered in this paper must be real.  In general, Landau damping does
not occur in modes of the Van Kampen type.  On the other hand,
perturbations which are superpositions of modes belonging to the
continuous spectrum would be expected to suffer Landau damping (Van Kampen
1955, 1957; Case 1959; see also Appendix 5.A. in Binney \& Tremaine 1987).
Our concentration on a normal mode analysis puts the study of
Landau damping beyond the scope of the present investigation.  The present
calculations give no evidence for the existence in the systems studied of
a discrete spectrum of radial modes with frequencies lying outside the
continuous spectrum.

In the N-body experiments on the fundamental mode of radial oscillation
that are represented in Figure 7(a), the orbits that are populated with
stars have frequencies of the radial motions which do not span the entire
range described by inequality (111).  These experiments were performed
on systems of 1000 bodies.  The density of stars is so low
in the regions of the phase space accessible to orbits with the lowest
frequencies that these N-body realizations of the systems studied do not
include such stars.  In other words, the N-body realizations of these
systems do not include stars that would resonate with the fundamental mode.
Thus the fundamental mode appears to be a discrete mode in the N-body
experiments, and that perturbation should not suffer Landau damping.  
For substantially larger (and more realistic) numbers of bodies, N-body
realizations would include stars that resonate with the frequency
of the fundamental mode, and perturbations that include the fundamental
mode could suffer Landau damping.  The damping would be expected to
be slow, because the density of resonant stars in the phase space
would be very low.

\section{Concluding remarks}

For the study of small perturbations in stellar systems, the matrix
method described in this paper has two particularly attractive features.
The first is that the characteristic value problem for modes in the
matrix representation is linear in the frequencies of the modes.
Consequently, the solution of the characteristic equation for the
frequencies is readily obtained with the aid of standard matrix
methods.  In contrast, the matrix method of Kalnajs involves a
characteristic equation which is nonlinear in the frequencies, and
it is a more challenging task to obtain the solutions for the
frequencies.  The second feature that makes the present matrix
method particularly attractive is that the underlying Lagrangian
representation deals directly with the perturbations of the
trajectories of the particles.  The comparison of analytic
investigations of small perturbations with N-body experiments
is straight forward in this case.  Analytic solutions for the
Lagrangian displacements of the perturbations can be used as
initial conditions for N-body experiments which provide direct
tests of analytic results.  The experimental results included in
Figure 7(a) represent such a test.

The results obtained in Section 6 consist of relatively accurate
representations of six modes at central concentrations of the order of
10, quite an accurate representation of the fundamental mode of radial
oscillation at central concentrations ranging up to the order of 10,000,
and an approximate delineation of three other modes at central
concentrations ranging up to the order of 100-1,000.  These results
were obtained with the aid of a very modest computational effort.
Therefore, considerable scope remains for more diverse investigations
of modes of oscillation and instability in stellar systems with
the aid of more extensive numerical calculations.

For these reasons, the planned extension of the present investigation
to nonradial modes in spheres and to modes in axisymmetric systems,
as described at the end of Section 1, would appear to hold considerable
promise.

\section*{Acknowledgments}

The research described in this paper was begun as a direct result of
work done by the author during a stay at the Leiden University
Observatory in the spring of 1998, and the early sections of this
paper were written during a second stay there in the spring of 2000.
I am grateful to George Miley and Tim de Zeeuw for the
opportunities to work in that wonderfully stimulating environment.
Tim de Zeeuw has been a source of much good advice and helpful
encouragement.  Conversations with Bernard F. Schutz, Christopher Hunter,
and Richard H. Miller have been a great help in the illumination and
clarification of technical difficulties.  Christopher Hunter, David Merritt,
Martin Weinberg, and Tim de Zeeuw generously took time and effort to
read an earlier version of this paper, and their remarks and suggestions
contributed substantially to the revisions leading to the present version.

\appendix

\section{List of vector polynomials}

In this appendix, we list additional groups of vector polynomials
identified with the aid of the procedure described at the beginning
of Section 4.2.

The vector polynomials in the case that $k=1$ are
$$
\eqalign{
&{\eqalign{
\bpi(1,1) &= \bmu_{+1}^{+1} (1,0,0) ,\cr
\bpi(1,3) &= \bmu_{+1}^{+1} (0,1,0) + {1 \over 2} \bmu_{-1}^{-1} (1,0,0)
- {1 \over 2} \bmu_{+1}^{-1} (1,0,0) ,\cr}}
\qquad
{\eqalign{
\bpi(1,2) &= \bmu_{-1}^{+1} (0,0,1) , \cr
\bpi(1,4) &= \bmu_{-1}^{+1} (0,1,0) + {1 \over 2} \bmu_{+1}^{-1} (0,0,1)
- {1 \over 2} \bmu_{-1}^{-1} (0,0,1), \quad \hbox{ and } \cr}} \cr
&\bpi(1,5) = \bmu_{+1}^{+1} (0,0,1) - \bmu_{-1}^{+1} (1,0,0)
+ 2 \bmu_{-1}^{-1} (0,1,0) - 2 \bmu_{+1}^{-1} (0,1,0) ,\cr}
 \eqno\stepeq
$$
where we suppress the arguments ${\bmath{z}}_{+1} $ and $ {\bmath{z}}_{-1} $
here and in what follows.

Likewise, the vector polynomials in the case that $k=2$ are
$$
\eqalign{
&{\eqalign{
\bpi(2,1) &= \bmu_{+1}^{+1} (2,0,0) ,\cr
\bpi(2,3) &= \bmu_{+1}^{+1} (1,1,0) + {1 \over 4} \bmu_{-1}^{-1} (2,0,0)
- {1 \over 4} \bmu_{+1}^{-1} (2,0,0) ,\cr}}
\qquad
{\eqalign{
\bpi(2,2) &= \bmu_{-1}^{+1} (0,0,2) , \cr
\bpi(2,4) &= \bmu_{-1}^{+1} (0,1,1) + {1 \over 4} \bmu_{+1}^{-1} (0,0,2)
- {1 \over 4} \bmu_{-1}^{-1} (0,0,2), \cr}} \cr
&\bpi(2,5) = \bmu_{+1}^{+1} (1,0,1) + 2 \bmu_{+1}^{+1} (0,2,0)
+ 2 \bmu_{-1}^{-1} (1,1,0) - { 1 \over 2 } \bmu_{-1}^{+1} (2,0,0)
- 2 \bmu_{+1}^{-1} (1,1,0) ,\cr
&\bpi(2,6) = \bmu_{-1}^{+1} (1,0,1) + 2 \bmu_{-1}^{+1} (0,2,0)
+ 2 \bmu_{+1}^{-1} (0,1,1) - { 1 \over 2 } \bmu_{+1}^{+1} (0,0,2)
- 2 \bmu_{-1}^{-1} (0,1,1) , \quad \hbox{ and } \cr
&\bpi(2,7) = \bmu_{+1}^{+1} (0,1,1) + { 1 \over 2 } \bmu_{-1}^{-1} (1,0,1)
+ \bmu_{-1}^{-1} (0,2,0)
- \bmu_{-1}^{+1} (1,1,0) - { 1 \over 2 } \bmu_{+1}^{-1} (1,0,1)
- \bmu_{+1}^{-1} (0,2,0) .\cr }
 \eqno\stepeq
$$

Finally, the vector polynomials in the case that $k=3$ are
$$
\eqalign{
&{\eqalign{
\bpi(3,1) &= \bmu_{+1}^{+1} (3,0,0) ,\cr
\bpi(3,3) &= \bmu_{+1}^{+1} (2,1,0) + {1 \over 6} \bmu_{-1}^{-1} (3,0,0)
- {1 \over 6} \bmu_{+1}^{-1} (3,0,0) ,\cr}}
\qquad
{\eqalign{
\bpi(3,2) &= \bmu_{-1}^{+1} (0,0,3) , \cr
\bpi(3,4) &= \bmu_{-1}^{+1} (0,1,2) + {1 \over 6} \bmu_{+1}^{-1} (0,0,3)
- {1 \over 6} \bmu_{-1}^{-1} (0,0,3), \cr}} \cr
&\bpi(3,5) = \bmu_{+1}^{+1} (2,0,1) + 4 \bmu_{+1}^{+1} (1,2,0)
+ 2 \bmu_{-1}^{-1} (2,1,0) - { 1 \over 3 } \bmu_{-1}^{+1} (3,0,0)
- 2 \bmu_{+1}^{-1} (2,1,0) ,\cr
&\bpi(3,6) = \bmu_{-1}^{+1} (1,0,2) + 4 \bmu_{-1}^{+1} (0,2,1)
+ 2 \bmu_{+1}^{-1} (0,1,2) - { 1 \over 3 } \bmu_{+1}^{+1} (0,0,3)
- 2 \bmu_{-1}^{-1} (0,1,2) ,\cr
&\bpi(3,7) = \bmu_{+1}^{+1} (1,1,1) + { 2 \over 3 } \bmu_{+1}^{+1} (0,3,0)
+ { 1 \over 4 } \bmu_{-1}^{-1} (2,0,1) + \bmu_{-1}^{-1} (1,2,0) 
- { 1 \over 2 } \bmu_{-1}^{+1} (2,1,0) - { 1 \over 4 } \bmu_{+1}^{-1} (2,0,1)
- \bmu_{+1}^{-1} (1,2,0) ,\cr
&\bpi(3,8) = \bmu_{-1}^{+1} (1,1,1) + { 2 \over 3 } \bmu_{-1}^{+1} (0,3,0)
+ { 1 \over 4 } \bmu_{+1}^{-1} (1,0,2) + \bmu_{+1}^{-1} (0,2,1) 
- { 1 \over 2 } \bmu_{+1}^{+1} (0,1,2) - { 1 \over 4 } \bmu_{-1}^{-1} (1,0,2)
- \bmu_{-1}^{-1} (0,2,1) , \cr
&\hbox{ and } \cr
&\eqalign{\bpi(3,9) = \bmu_{+1}^{+1} (1,0,2)  + 4 \bmu_{+1}^{+1} (0,2,1)
&+ 4 \bmu_{-1}^{-1} (1,1,1) + { 8 \over 3 } \bmu_{-1}^{-1} (0,3,0) \cr
&- \bmu_{-1}^{+1} (2,0,1)  - 4 \bmu_{-1}^{+1} (1,2,0)
- 4 \bmu_{+1}^{-1} (1,1,1) - { 8 \over 3 } \bmu_{+1}^{-1} (0,3,0) . \cr }
\cr }
 \eqno\stepeq
$$

\section{Inner products of vector polynomials and their adjoints}

The construction of the basis vectors described in Section 4.3
requires the evaluation of the inner products
${\langle A \bpi (q^{\prime} ) , \bpi (q) \rangle}_2$.
These can be expressed in terms of the inner products
$\langle A \bmu_{\sigma^\prime}^{\tau^\prime} ( m^\prime, n^\prime,
p^\prime ) , \bmu_\sigma^\tau (m,n,p) \rangle_2$.

Making use of equations
(42) and (50) for the vector monomials and their adjoint vectors,
respectively, constructing the inner products of those vectors in
accordance with equation (9), and taking the complex conjugates
of quantities in accordance with equations (49), we obtain
$$
\langle A \bmu_{\sigma^\prime}^{\tau^\prime} ( m^\prime, n^\prime, p^\prime ) ,
\bmu_\sigma^\tau (m,n,p) \rangle_2
= {\rm i}n_0 ({\tau^\prime} {\sigma^\prime} + \tau \sigma) m_* \int_\Omega
\bmath{z}_{- {\sigma^\prime}} \cdot \bmath{z}_\sigma
M ( m+p^\prime , n+n^\prime , p+m^\prime ) f_0
{\rm d} \bmath{x} {\rm d} \bmath{v} .
\eqno\stepeq
$$

By inspection of equation (B1), we see that
$$
\langle A \bmu_{\sigma^\prime}^{\tau^\prime} ( m^\prime, n^\prime, p^\prime ) ,
\bmu_\sigma^\tau (m,n,p) \rangle_2 = 0 \quad \hbox { unless }
{\tau^\prime} {\sigma^\prime} = \tau \sigma . \eqno\stepeq
$$
We also note that the quantity $ \bmath{z}_{- {\sigma^\prime}} \cdot
\bmath{z}_\sigma M ( m+p^\prime , n+n^\prime , p+m^\prime ) $ can be
expressed as a polynomial in the components of $\bmath{x}$ and
$ {\rm i} \bmath{v} $ in virtue of equations (33) and (35).  For
systems of the kind considered in Section 4, in
which $f_0 (\bmath{x},\bmath{v})$ is an even function of $\bmath{v}$,
it follows that the imaginary part of the integrand on the
right-hand side of equation (B1) is an odd function of the components
of $\bmath{v}$.  Thus the integral on the right-hand side of equation (B1)
is a real quantity, and we have the result that
$ \langle A \bmu_{\sigma^\prime}^{\tau^\prime} ( m^\prime, n^\prime, p^\prime ) ,
\bmu_\sigma^\tau (m,n,p) \rangle_2 $ is an imaginary quantity for systems
in which the distribution function $f_0$ is an even function of the
components of the velocity.  Finally, we verify that
$$
\langle A \bmu_\sigma^\tau (m,n,p) ,
\bmu_{\sigma^\prime}^{\tau^\prime}
( m^\prime, n^\prime, p^\prime ) \rangle_2 =
\langle A \bmu_{\sigma^\prime}^{\tau^\prime} ( m^\prime, n^\prime, p^\prime ) ,
\bmu_\sigma^\tau (m,n,p) \rangle_2 .
\eqno\stepeq
$$
For, as we have just explained, the integral on the right-hand side
of equation (B1) is real.  Therefore, we may replace the integrand there
with its complex conjugate, rewrite the result with the aid of the first
two of equations (49) and, in virtue of equation (B1), transform an
expression for the right-hand side of equation (B3) into an expression
for the left-hand side of equation (B3).

It follows from equations (55) and (56) that
$$
{\langle A \bpi (q^{\prime} ) , \bpi (q) \rangle}_2 = 
\sum_{j^{\prime} = 1}^{J^{\prime}} \sum_{j=1}^J
\langle {q^{\prime}}| {j^{\prime}} \rangle \langle q|j \rangle
{ \langle A \bmu_{\sigma^{\prime}}^{\tau^{\prime}} (m^{\prime},n^{\prime}
,p^{\prime}) , \bmu_{\sigma}^{\tau} (m,n,p) \rangle }_2 ,
\eqno\stepeq
$$
where the values
$J=J (q)$,
$\sigma = \sigma (q;j)$,
$\tau = \tau (q;j)$,
$m=m (q;j)$
 $n=n (q;j)$,
$p=p (q;j)$,
$J^{\prime}=J (q^{\prime})$,
$\sigma^{\prime} = \sigma (q^{\prime};j^{\prime})$,
$\tau^{\prime} = \tau (q^{\prime};j^{\prime})$,
$m^{\prime}=m (q^{\prime};j^{\prime})$
 $n^{\prime}=n (q^{\prime};j^{\prime})$,
and $p^{\prime}=p (q^{\prime};j^{\prime})$ are determined by inspection
of equations (45) and (A1)-(A3), as we have explained following equation (55).

The inner products ${\langle A \bpi (q^{\prime} ) , \bpi (q) \rangle}_2$
have the following properties.

\beginlist
\item (i) Inasmuch as the coefficients
$\langle {q^{\prime}}| {j^{\prime}} \rangle$ and $\langle q|j \rangle$
are real constants in equation (B4) and the inner products
$\langle A \bmu_{\sigma^\prime}^{\tau^\prime} ( m^\prime, n^\prime,
p^\prime ) , \bmu_\sigma^\tau (m,n,p) \rangle_2$ are imaginary
quantities, the inner products
${\langle A \bpi (q^{\prime} ) , \bpi (q) \rangle}_2$ are
imaginary quantities.
\item (ii) Rewriting equation (B4) with the aid of equation (B3),
we readily verify that
${\langle A \bpi (q^{\prime} ) , \bpi (q) \rangle}_2$
has the symmetry
$$
{ \langle A \bpi (q^\prime ) , \bpi (q) \rangle }_2
= { \langle A \bpi (q) , \bpi (q^\prime ) \rangle }_2 \qquad
(q^\prime =1,2,\dots,N;q=1,2,\dots,N) . \eqno\stepeq
$$
\item (iii) A consequence of the first of equations (54) is that
$$
\eqalign{
{ \langle A \bpi (q^\prime ) , \bpi (q) \rangle }_2
= { \langle A \bpi (q^\prime -1), \bpi (q-1) \rangle }_2^* 
\quad &\hbox { and } \quad 
{ \langle A \bpi (q^\prime ) , \bpi (q-1) \rangle }_2
= { \langle A \bpi (q^\prime -1), \bpi (q) \rangle }_2^* \cr
(q^\prime =2,4,\dots,Q;&q=2,4,\dots,Q) , \cr } \eqno\stepeq
$$
inasmuch as complex conjugation commutes with the operator $A$.
\item (iv) Likewise, it follows from the second of equations (54) that
$$
{ \langle A \bpi (q^\prime ) , \bpi (q) \rangle }_2 = 0 \qquad
(q^\prime = Q +1,Q +2, \dots,N;
q=Q +1,Q +2, \dots,N) . \eqno\stepeq
$$
More specifically, the second of equations (54) implies that the inner
product on the left-hand side of equation (B7) is equal to its complex
conjugate and is therefore real.  However, the general result described
in item (i) above implies that this inner product is imaginary.  Therefore,
the inner product described in equation (B7) must vanish.  Equation (B7)
is essentially a set of orthogonality relations for the purely imaginary
vector polynomials and their adjoint vectors.  In the case $ q^\prime = q $,
we see that the inner product of a purely imaginary vector polynomial and
its own adjoint vector vanishes.  In other words, the purely imaginary
vector polynomials are null vectors.  It is for this reason
that some of the basis
vectors constructed in Section 4.3 turn out to be null vectors
as well.
\item (v) A further consequence of equations (54) is that
$$
{ \langle A \bpi (q^\prime ) , \bpi (q) \rangle }_2 = 
{ \langle A \bpi (q^\prime ) , \bpi (q-1) \rangle }_2 \qquad
(q^\prime = Q +1,Q +2, \dots,N;q=2,4,\dots,Q) . \eqno\stepeq
$$
For, when we substitute from equations (54) for the vector polynomials
on the left-hand side of equation (B8) and recall that complex conjugation
and $A$ are commuting operators and that these inner products are
imaginary quantities, we obtain the inner product on the right-hand
side of equation (B8).
\item (vi) Finally, we observe that
$$
{ \langle A \bpi (q) , \bpi (q-1) \rangle }_2 = 
{ \langle A \bpi (q-1) , \bpi (q) \rangle }_2 = 0 \qquad
( q=2,4,\dots,Q) . \eqno\stepeq
$$
The first of the equalities in equation (B9) is a special case of
equation (B5).  When we combine that symmetry relation
with the result of putting $q^\prime = q$
in the second of equations (B6), we find that the inner products
in equation (B9) must be real.  On the other hand, such inner
products have been shown to be imaginary, in general, so they
must vanish as claimed in equation (B9).  Equation (B9) is
essentially a set of orthogonality relations for the complex-conjugate
pairs of vector polynomials and their adjoint vectors.
\endlist

Our procedure for the evaluation of the integral over the phase space
on the right-hand side of equation (B1) is described in Appendix E
below.  The numerical values of the inner products
${\langle A \bpi (q^{\prime} ) , \bpi (q) \rangle}_2$ derived with the
aid of equation (B4) have been checked with the aid of equations (B5)-(B9).

\section{Certain identities}

The arrangement of the basis vectors $\bbeta (q)$ $(0 < q \le Q)$
in complex-conjugate pairs as described in Section 4.3.1 (see eq. [68]) is a
consequence of the identities presented in equation (69).
In what follows, we shall prove equations (69) by induction.

We begin the proof by reducing the second of equations (65) and
equation (67) in the case that $q=2$ to
$$
c(2,1) = - N(1) { \langle A \bpi (1), \bpi (2) \rangle }_2 = 0
\quad \hbox {and} \quad
N^2 (2) = { \left[ { \langle A \bpi (2), \bpi (2) \rangle }_2
- c^2 (2,1) \right] }^{-1}
= [ { \langle A \bpi (1), \bpi (1) \rangle }_2^* ]^{-1}
= [N^2(1)]^* , \eqno\stepeq
$$
respectively, with the aid of equations (54) and (B9) and the first
of equations (65).  It follows from the first of equations (C1) that
we can reduce equation (66) in the case that $s=2$ to
$$
c(q,2)
= N(2) \left[ c(2,1)c(q,1) 
- { \langle A \bpi (2), \bpi (q) \rangle }_2 \right]
= - N(2) { \langle A \bpi (2), \bpi (q) \rangle }_2
\qquad (q = 3,4,\dots,Q ) .
\eqno\stepeq
$$
Equations (B6) and the second of equations (C1) (see also eqs. [65])
enable us to derive the relations
$$
\eqalign{
c(q,2) = -N^* (1) { { \langle A \bpi(1) , \bpi (q-1) \rangle }_2 }^*
= c^* (q-1,1) \quad &\hbox {and} \quad
c(q,1) = -N^* (2) { { \langle A \bpi(2) , \bpi (q-1) \rangle }_2 }^*
= c^* (q-1,2) \cr
(q = 4,6,&\dots,Q ) \cr }
\eqno\stepeq
$$    
from equation (C2).  Equations (C1) establish the first two of equations (69)
in the case that $q=2$, and equations (C3) establish the third and fourth of
equations (69) in the case that $q = 4,6,\dots,Q$ and $s=2$.

Continuing the proof of equations (69), we now consider equation (66)
for given, even values of $q$ and $s$, we recall that the vector
polynomials satisfy equations (B6), and we assume, in the sense of a
proof by induction, that $N(s)$, $c(s,r)$, and $c(q,r)$ $(r<s<q)$
satisfy equations (69).  Under such conditions, we can rewrite
equation (66) in the manner
$$
\eqalign{ c(q,s)
&= N^* (s-1) \left[ \sum_{r=1}^{s-2} c^* (s-1,r)c^* (q-1,r) 
- { \langle A \bpi (s-1), \bpi (q-1) \rangle }_2^* \right] \cr
&= c^* (q-1,s-1) \qquad (q=6,8,\dots,Q;s=4,6,\dots,q-2) . \cr }
\eqno\stepeq
$$
In this reduction, we break the sum on the right-hand side of equation (66)
into sums over even and odd values of $r$, rewrite the separate sums with
the aid of the third and fourth of equations (69), suppress $c(q,q-1)$ in
accordance with the second of equations (69), and recombine the
results in the single sum shown in equation (C4).  In the
verification of this derivation, it is helpful to note that
the application of the third and fourth of equations (69) turns
sums over even and odd values of $r$ into sums over odd and even
values of $r-1$, respectively.  Under the same conditions and with
similar reductions, we can also rewrite equation (66) in the manner
$$
\eqalign{ c(q,s-1)
&= N^* (s) \left[ \sum_{r=1}^{s-1} c^* (s,r)c^* (q-1,r) 
- { \langle A \bpi (s), \bpi (q-1) \rangle }_2^* \right] \cr
&= c^*(q-1,s) \qquad (q=6,8,\dots,Q;s=4,6,\dots,q-2) \cr }
\eqno\stepeq
$$
and, in the case that $s = q-1$, in the manner
$$
\eqalign{ c(q,q-1)
&= N(q-1) \sum_{r=1}^{q-2} c(q-1,r)c(q,r) \cr
&= N(q-1) \sum_{r=1}^{q-2} c^* (q,r)c^* (q-1,r) = 0
\qquad (q=4,6,\dots,Q) . \cr }
\eqno\stepeq
$$
The quantity $c(q,q-1)$ must vanish here, because, the equality of the
two sums in equation (C6) implies, on the one hand that each sum is a
real quantity, whereas our discussion just prior to Section 4.3.1
shows, on the other hand, that the terms $c(q-1,r)c(q,r)$ and
$c^* (q,r)c^* (q-1,r)$ in those sums must be imaginary quantities.
Finally, we can transform the right-hand side of equation (67) with
the aid of equations (B6) and (69) and derive
$$
N^2 (q) = { \biggl( { \langle A \bpi (q-1), \bpi (q-1) \rangle }_2^*
- \sum_{r=1}^{q-2}  [c^2 (q-1,r)]^* \biggr) }^{-1} = [N^2 (q-1)]^*
\qquad (q=4,6,\dots,Q) . \eqno\stepeq
$$

The completion of the proof of equations (69) now consists of starting
with the results in equations (C1) and (C3), considering equations (C4)-(C7)
in the sequence $q=3,4,\dots,Q$, and, for
each value of $q$, verifying equations (C4) and (C5) in the sequence
$s=2,3,\dots,q-2$ and then verifying equations (C6) and (C7).  This
confirms the arrangement of the basis vectors $\bbeta (q)$ $(0 < q \le Q)$
in complex-conjugate pairs as described in equation (68).

The construction of the null basis vectors in Section 4.3.2 requires the reduction of a
subset of equations (61) and (62) to equations (72)-(74).  We conclude this
appendix with the derivation of identities that are required for that
reduction.  We must first prove that the
quantities $c(q,s)$ which satisfy equations (70) and (71) also satisfy
the identities
$$
c(q,s) = - c^* (q,s-1) \qquad (q=Q+1,Q+2,\dots,N ; s=2,4,\dots,Q) .
\eqno\stepeq
$$

For a proof of equation (C8) by induction, we first consider equation (71)
in the case that $s=2$, and we make use of equations (54), (C1), and (70)
in order to write
$$
c(q,2) = - N(2) { \langle A \bpi (2), \bpi (q) \rangle }_2
= N^* (1) { \langle A \bpi (1), \bpi (q) \rangle }_2^*
= - c^* (q,1) \qquad (q=Q+1,Q+2,\dots,N) , \eqno\stepeq
$$
which proves the identity in the case that $s=2$.
In order to complete the proof, we must show that the condition
$c(q,r) = - c^* (q,r-1)$ holds for $r=s$, where $s(>2)$ is an even integer,
if that condition holds for even values of $r$ which are less than $s$.
For this purpose, we make use of equations (54) and (69) in order
to reduce equation (71) to
$$
\eqalign{ c(q,s)
&= - N^* (s-1) \left[ \sum_{r=1}^{s-2} c^* (s-1,r)c^* (q,r) 
- { \langle A \bpi (s-1), \bpi (q) \rangle }_2^* \right] \cr
&= - c^* (q,s-1)
\qquad (q=Q+1,Q+2,\dots,N ; s=4,6,\dots,Q) . \cr }
\eqno\stepeq
$$
In this reduction, we break the sum on the right-hand side of equation (71)
into sums over even and odd values of $r$, rewrite the separate sums with
the aid of the third and fourth of equations (69) and the condition
$c(q,r) = - c^* (q,r-1) \quad (r=2,4,\dots,s-2)$, make use of the second
of equations (69) in order to suppress $c(s,s-1)$, and recombine the
results in the single sum shown in equation (C10).  That the transformed
expression for $c(q,s)$ in equation (C10) is equal to $-c^* (q,s-1)$ is also
a consequence of equation (71).

The identity that is required for the derivation of equations (72)-(74) is
$$
\sum_{r=1}^{Q} c(s,r)c(q,r) = \sum_{r=1}^{Q} c^* (s,r)c^* (q,r) = 0
\qquad (q=Q+2,Q+3,\dots,N;s=Q+1,Q+2,\dots,q) . \eqno\stepeq
$$
In this reduction, we break the first sum in equation (C11)
into sums over even and odd values of $r$, rewrite the separate sums with
the aid of equations (C8), and recombine the
results in the second sum shown in equation (C11).
The two sums in equation (C11) must vanish, because, their equality
implies, on the one hand, that each sum is a
real quantity, whereas our earlier discussion of equations (61)-(63)
shows, on the other hand, that the terms $c(s,r)c(q,r)$ and
$c^* (s,r)c^* (q,r)$ in those sums must be imaginary quantities.

\section{The matrix of the operator $P$}

In order to evaluate the matrix
${ \langle A \bbeta (q^{\prime}), P \bbeta (q) \rangle }_2$,
we apply the operator $P$ to the right-hand sides of equations (57),
and we form the inner product of $A \bbeta (q^{\prime})$ with the
resulting expressions.  We obtain
$$
{ \langle A \bbeta (q^{\prime}), P \bbeta (1) \rangle }_2
= N(1)  { \langle A \bbeta (q^{\prime}), P \bpi (1) \rangle }_2
\eqno\stepeq
$$
and
$$
{ \langle A \bbeta (q^{\prime}), P \bbeta (q) \rangle }_2
= N(q) \left[ \sum_{r=1}^{q-1} c(q,r)
{ \langle A \bbeta (q^{\prime}), P \bbeta (r) \rangle }_2
+ { \langle A \bbeta (q^{\prime}), P \bpi (q) \rangle }_2 \right]
\qquad (q > 1) .
\eqno\stepeq
$$
In order to evaluate the right-hand sides of equations
(D1) and (D2), we form the inner products of the adjoint vectors
$ A \bbeta (q^{\prime})$, expressed as in equations (59), with
the quantities $P \bpi (q)$.  We obtain
$$
{ \langle A \bbeta (1), P \bpi (q) \rangle }_2
= N(1) { \langle A \bpi (1), P \bpi (q) \rangle }_2
\eqno\stepeq
$$
and
$$
{ \langle A \bbeta (q^{\prime}), P \bpi (q) \rangle }_2
= N(q^{\prime}) \left[ \sum_{r^{\prime}=1}^{q^{\prime}-1}
c(q^{\prime},r^{\prime})
{ \langle A \bbeta (r^{\prime}), P \bpi (q) \rangle }_2
+ { \langle A \bpi (q^{\prime}), P \bpi (q) \rangle }_2 \right]
\qquad (q^{\prime} > 1) .
\eqno\stepeq
$$
Finally, we construct the matrix
$$
{ \langle A \bpi (q^{\prime} ) , P \bpi (q) \rangle }_2 = 
\sum_{j^{\prime} = 1}^{J^{\prime}} \sum_{j=1}^J
\langle {q^{\prime}}| {j^{\prime}} \rangle \langle q|j \rangle
{ \langle A \bmu_{\sigma^{\prime}}^{\tau^{\prime}} (m^{\prime},n^{\prime}
,p^{\prime}) , P \bmu_{\sigma}^{\tau} (m,n,p) \rangle }_2
\eqno\stepeq
$$
(see eqs. [55] and [B4]) in order to evaluate the
right-hand sides of equations (D3) and (D4).

Evidently, the calculations involved in the evaluation of
${ \langle A \bbeta (q^{\prime}), P \bbeta (1) \rangle }_2$
begin with the evaluation of the matrix
${ \langle A \bmu_{\sigma^{\prime}}^{\tau^{\prime}} (m^{\prime},n^{\prime}
,p^{\prime}) , P \bmu_{\sigma}^{\tau} (m,n,p) \rangle }_2$ and continue
with the evaluation of
${ \langle A \bpi (q^{\prime} ) , P \bpi (q) \rangle }_2$.  We next consider
equations (D3) and (D4) in the sequence $q=1,2,\dots,$ and, for each
value of $q$ in turn, evaluate the quantities 
${ \langle A \bbeta (q^{\prime}), P \bpi (q) \rangle }_2$ in the
sequence $q^{\prime}=1,2,\dots$.  At each step, the right-hand side of
equation (D3) or (D4) can be evaluated in terms of quantities whose
values are known results of preceding steps in the procedure.
Likewise, we finally consider
equations (D1) and (D2) in the sequence $q=1,2,\dots,$ and, for each
value of $q$ in turn, evaluate the quantities 
${ \langle A \bbeta (q^{\prime}), P \bbeta (q) \rangle }_2$ in the
sequence $q^{\prime}=1,2,\dots$.  Similarly, the right-hand side of
equation (D1) or (D2) can be evaluated at each step in terms of
quantities whose values are known results of preceding steps. 

We have now reduced numerical work required for the construction of the
matrix ${ \langle A \bbeta (q^{\prime}), P \bbeta (q) \rangle }_2$
to the evaluation of the matrix
$$
\eqalign{
{ \langle A \bmu_{\sigma^{\prime}}^{\tau^{\prime}} (m^{\prime},n^{\prime}
,p^{\prime}) , P \bmu_{\sigma}^{\tau} (m,n,p) \rangle }_2 =
&-{\rm i}{\tau^{\prime}}{\sigma^{\prime}} \tau \sigma {n_0}^2 m_*
\int_\Omega \bmath{z}_{- {\sigma^\prime}} \cdot \bmath{z}_\sigma
M ( m+p^\prime , n+n^\prime , p+m^\prime ) f_0
{\rm d} \bmath{x} {\rm d} \bmath{v} \cr
&-{\rm i}({\tau^{\prime}}{\sigma^{\prime}} + \tau \sigma ) n_0 m_*
\int_\Omega M ( p^\prime , n^\prime , m^\prime )
\bmath{z}_{- {\sigma^\prime}} \cdot D [ \bmath{z}_\sigma
M ( m , n , p ) ] f_0
{\rm d} \bmath{x} {\rm d} \bmath{v} \cr
&+{\rm i} m_* \int_\Omega M ( p^\prime , n^\prime , m^\prime )
\bmath{z}_{- {\sigma^\prime}} \cdot \Delta \bmath {a} \{ \bmath{z}_\sigma
M ( m , n , p ) \} f_0
{\rm d} \bmath{x} {\rm d} \bmath{v} . \cr }
\eqno\stepeq
$$
In the derivation of equation (D6), we have constructed
${ \langle A \bmu_{\sigma^{\prime}}^{\tau^{\prime}} (m^{\prime},n^{\prime}
,p^{\prime}) , P \bmu_{\sigma}^{\tau} (m,n,p) \rangle }_2$ in accordance
with the definition of an inner product given in equation (9),
we have substituted from equations (50), (6), and (42) for 
$A \bmu_{\sigma^{\prime}}^{\tau^{\prime}} (m^{\prime},n^{\prime}
,p^{\prime})$, $P$, and $\bmu_{\sigma}^{\tau} (m,n,p)$, respectively,
in the integral to be evaluated, and we have reduced equation (D6)
to the form given here with the aid of equations (35) and (49).

The evaluation of the third integral on the right-hand side of
equation (D6) requires the construction of the quantity
$\Delta \bmath{a} \{ \bmath{z}_\sigma M ( m , n , p ) \}$.
Let $\rho_0 (r)$, $M_0 (r)$, and $V_0 (r)$ denote the density, the mass
interior to a spherical surface of radius $r$, and the gravitational
potential, respectively in the unperturbed system.  By definition and
by Newton's theorem, we have
$$
M_0 (r) = 4 \upi \int_0^r \rho_0 (r) r^2 {\rm d}r
\quad \hbox{and} \quad
- { { \upartial V_0 } \over { \upartial \bmath{x}}} =
- { {G M_0 (r) \bmath{x} } \over {r^3} } , \eqno\stepeq
$$
respectively.

For the radial oscillations of a sphere, it is convenient to construct
the Eulerian perturbation of the gravitational acceleration in terms
of the Eulerian perturbation of the density $\rho_1 (\bmath{x},t)$ and
the hydrodynamical Lagrangian displacement $\bxi (\bmath{x},t)$
(see equations [109]).  The perturbation $\rho_1$ is spherically symmetric, so,
by Newton's theorem and then Gauss' theorem, the Eulerian perturbation of
the gravitational acceleration reduces to
$$
- { { \upartial V_1 } \over { \upartial \bmath{x}}} 
= { { \bmath{x} } \over {r^3} } G \int_{|\bmath{x}|<r}
\left[ {\upartial \over {\upartial \bmath{x} } } \cdot
(\rho_0 \bxi ) \right] {\rm d} \bmath{x}
= {\bmath{x} \over {r^3}} 4 \upi r^2 G \rho_0 (r) {  \bmath{x}  \over r } \cdot
\bxi (\bmath{x},t) = 
4 \upi Gm_* \int \Delta \bmath{x} (\bmath{x},\bmath{v},t) f_0 (\bmath{x},\bmath{v})
{\rm d} \bmath{v} . \eqno\stepeq
$$
In the reduction of equation (D8), we have noted that $\bxi$ lies in
the radial direction under conditions of spherical symmetry, we have
identified $\bmath{x}/r$ as a unit vector in the radial direction,
and we have substituted from the second of equations (109)
for $\rho_0 \bxi$.

Starting with the definition of the operator $\Delta \bmath{a} \{ \}$ in
equation (3), we insert the Eulerian perturbation of the gravitational
potential in accordance with equation (29), we make use of equations (D7)
and (D8) for $- { { \upartial V_0 } / { \upartial \bmath{x}}}$
and $- { { \upartial V_1 } / { \upartial \bmath{x}}}$, respectively,
and we find, after additional reductions
which are straight-forward, that
$$
\Delta \bmath{a} \{ \bmath{z}_\sigma M ( m , n , p ) \} =
-{ {G M_0 (r)} \over r^3 } \bmath{z}_\sigma M(m,n,p)
- { \bmath{x} \over r } \left[ { \upartial \over { \upartial r } }
\left( { {G M_0 (r)} \over r^3 } \right) \right]
\bmath{x} \cdot \bmath{z}_\sigma M(m,n,p)
+ 4 \upi G m_* \int \bmath{z}_\sigma M(m,n,p) f_0 {\rm d} \bmath{v} .
\eqno\stepeq
$$

\section{Integrals over the phase space}

The applications of the matrix method described in Sections 5 and 6
of this paper are limited to systems in which the unperturbed distribution
function is a function only of the energy of a star and the velocity
distribution is accordingly isotropic.  Consider in that case the
integrals over the velocity space on the right-hand side of
equation (D9) and within the integrals over the phase space on the
right-hand sides of equations (B1) and (D6).  The integrands can
be expressed as polynomials in the components of $\bmath{v}$ in virtue
of equations (7), (33), and (35).  Inasmuch as the unperturbed velocity
distribution is isotropic, integrations over direction in the velocity
space are readily performed, and the integrations over the velocity
space reduce to the evaluation of the even moments
$$
G(r,\mu ) = m_* \int {| \bmath{v} |}^\mu f_0 (\bmath{x},\bmath{v})
{\rm d} \bmath{v} \quad (\mu=0,2,\dots ) \eqno\stepeq
$$
of the velocity distribution.  It is a consequence of the collisionless
Boltzmann equation that the moments satisfy the hierarchy of ordinary
differential equations
$$
{{\rm d} \over {\rm d}r} G(r,\mu ) = -(\mu + 1) G(r,\mu -2 )
{{{\rm d} V_0 } \over {{\rm d} r}} \quad (\mu = 2,4,\dots.) .
\eqno\stepeq
$$

Thus, in the evaluation of the integrals on the right-hand sides of
equations (B1) and (D6), we calculate the moments $G(r,\mu )$ of the
velocity distribution by integrating equations (E2) with the boundary
conditions $G(r,\mu ) = 0$ at $r = R$, where $R$ is the radius of
the unperturbed system.  We note in this connection that the zeroth
moment of the velocity distribution is
$G(r,0) = m_* \int f_0 (\bmath{x},\bmath{v}) {\rm d}\bmath{v}
= \rho_0 (r)$.

The evaluation of the moments of the velocity distribution
reduces the integrands on the right-hand sides of equations
(B1) and (D6) to functions of the radial coordinate $r= | \bmath{x} |$.
Accordingly, these integrations over the configuration space reduce to
quadratures over the radial coordinate $r$.  Those quadratures can be
reformulated in terms of the integration of associated ordinary differential
equations with appropriate boundary conditions.  The integration of
these equations can be carried out simultaneously with the integration
of equations (E2) described above.  In this work, those integrations
have been carried out with the aid of a standard Runge-Kutta integrator
of the fourth order (Press et al. 1986).


\section*{References}

\beginrefs
\bibitem 
\bibitem Antonov V. A., 1961, SvA, 4, 859
\bibitem 
\bibitem Antonov V. A., 1973, in Omarov G. B., ed., The Dynamics of
Galaxies and Star Clusters (Alma Ata: Nauka) (English translation
in de Zeeuw T., ed., 1987, Proc. IAU Symp. 127, Structure and Dynamics
of Elliptical Galaxies, Reidel, Dordrecht, p. 531)
\bibitem 
\bibitem Binney J., Tremaine S., 1987, Galactic Dynamics, Princeton
Univ. Press, Princeton
\bibitem
\bibitem Bernstein, I. B., Greene, J. M., Kruskal, M. D., 1957,
Phys. Rev., 108, 546.
\bibitem 
\bibitem Bertin G., Pegoraro F., Rubini F., Vesperini E., 1994,
ApJ, 434, 94
\bibitem 
\bibitem Bohm, D., Gross, E. P., 1949, Phys. Rev., 75, 1851.
\bibitem 
\bibitem Case K. M., 1959, Ann. Phys., 7, 349
\bibitem 
\bibitem 
\bibitem Fridman A. M., Polyachenko V. L., 1984, Physics of
Gravitating Systems, Springer-Verlag, New York
\bibitem 
\bibitem Jackson, J. D., 1960, Journal of Nuclear Energy Part C, 1, 171.
\bibitem 
\bibitem Kalnajs A. J., 1977, ApJ, 212, 637
\bibitem 
\bibitem Mathur S. D., 1990, MNRAS, 243, 529
\bibitem 
\bibitem Merritt D., 1987, in de Zeeuw T., ed., Proc. IAU Symp. 127,
Structure and Dynamics of Elliptical Galaxies, Reidel, Dordrecht, p. 315
\bibitem Merritt D., 1990, Proc. N. Y. Acad. Sci., 596, 150
\bibitem Merritt D., Sellwood J. A., 1994, ApJ, 425, 551
\bibitem 
\bibitem 
\bibitem 
\bibitem 
\bibitem Ostriker J. P., Peebles P. J. E., 1973, ApJ, 186, 287
\bibitem 
\bibitem Palmer P. L., 1994, Stability of Collisionless Stellar Systems:
Mechanisms for the Dynamical Structure of Galaxies, Kluwer Academic
Publishers, Dordrecht
\bibitem 
\bibitem Palmer P. L., Papaloizou J., 1987, MNRAS, 224, 1043
\bibitem 
\bibitem Polyachenko V. L., 1981, SvA Lett., 7, 79
\bibitem Polyachenko V. L., 1987, in de Zeeuw T., ed., Proc. IAU Symp. 127,
Structure and Dynamics of Elliptical Galaxies, Reidel, Dordrecht, p. 301
 
\bibitem Polyachenko V. L., Shukhman I. G., 1981, SvA, 25, 533
\bibitem Press W. H., Flannery B. P., Teukolsky S. A., Vetterling W. T.,
1986, Numerical Recipes, Cambridge University Press, Cambridge
\bibitem 
\bibitem Robijn F., 1995, PhD dissertation, Leiden Univ.
\bibitem 
\bibitem Saha P., 1991, MNRAS, 248, 494
\bibitem Sellwood J. A., Merritt D., 1994, ApJ, 425, 530
\bibitem Sellwood J. A., Valluri M., 1997, MNRAS, 287, 124
\bibitem 
\bibitem Toomre A. 1964, ApJ, 139, 1217
\bibitem 
\bibitem Vandervoort P, O., 1983, ApJ, 273, 511
\bibitem Vandervoort P, O., 1989, ApJ, 341, 105 (LM)
\bibitem Vandervoort P, O., 1991, ApJ, 377, 49
\bibitem Vandervoort P, O., 1999, MNRAS, 303, 393
\bibitem Vandervoort P, O., 2001, in preparation
\bibitem 
\bibitem van Kampen N. G., 1955, Physica, 21, 949
\bibitem van Kampen N. G., 1957, Physica, 23, 647
\bibitem 
\bibitem Vlasov A. A., 1945, J. Phys. (USSR) 9, 25
\bibitem 
\bibitem 
\bibitem Weinberg M. D., 1989, MNRAS, 239, 549
\bibitem Weinberg M. D., 1991a, ApJ, 368, 66 
\bibitem Weinberg M. D., 1991b, ApJ, 373, 391 
\bibitem 
\endrefs
\bye